\documentclass{article}
\usepackage[a4paper,top=3cm,bottom=3cm,left=2.1cm,right=2.1cm,marginparwidth=1.75cm]{geometry}

\usepackage{graphicx} 

\usepackage{bm}
\usepackage{amsmath, amsfonts, amssymb, amsthm}
\usepackage{mathrsfs, dsfont}
\usepackage[dvipsnames]{xcolor}
\usepackage{graphicx}
\usepackage{verbatim}
\usepackage{float}
\usepackage{comment}
\usepackage{tikz}
\usepackage{ulem}
\usepackage{subfig}
\usepackage{bigints}
\usetikzlibrary{positioning}

\usepackage[colorlinks=true, allcolors=blue]{hyperref}
\usepackage{xparse} 
\usepackage[TS1,T1]{fontenc}
\usepackage{tasks}
\usepackage{siunitx}
\usepackage[table]{xcolor}
\usepackage{colortbl}
\settasks{style=itemize}

\usepackage[authoryear]{natbib}

\newtheorem{theorem}{Theorem}
\newtheorem{proposition}{Proposition}

\newtheorem{example}{Example}

\usepackage{pifont}
\usepackage{enumitem}
\usepackage{mathtools}
\usepackage{bbm}
\usepackage{authblk}
\usepackage{shuffle}
\usepackage{multirow} 
\usepackage{booktabs} 

\newcommand{\emptyword}{{\textup{\textbf{\o{}}}}}
\newcommand{\word}[1]{{{\boldsymbol{#1}}}}

\def\layersep{2.5cm}

\newcommand{\corners}{6pt}

\title{Approximation of stochastic insurer balance-sheet results using signatures of economic scenarios}

\author[1]{Hervé Andrès}
\author[2]{Alexandre Boumezoued}
\author[3]{Arthur Bourdon\footnote{Corresponding author : arthur.bourdon@milliman.com; arthur.bourdon@enpc.fr}}
\author[4]{Benjamin Jourdain}

\affil[1,2,3]{Milliman R\&D}
\affil[3,4]{CERMICS, ENPC, Institut Polytechnique de Paris, CNRS, Marne-la-Vallée, France \& MATHRISK team-project, Inria Paris, France}
\date{\today}
\begin{document}

\maketitle

\begin{center}
\textbf{\large Abstract}
\end{center}

\setlength{\leftskip}{1cm} 
\setlength{\rightskip}{1cm} 
In the insurance industry, Asset and Liability Management (ALM) models are key tools for numerous applications, including Solvency Capital Requirement (SCR) computation and asset allocation optimization. However, their use often entails a significant computational cost, especially when a large number of sensitivities or stressed balance-sheet evaluations must be performed. In this work, we propose an approximation framework for the outputs of an ALM model, such as the Value In Force or the Best Estimate, based on the theory of path signatures. More precisely, the proposed approach consists of approximating ALM outputs by a linear combination of signature terms derived from input economic scenarios. We show that the resulting surrogate is easy to calibrate, essentially through regularized linear regression, and exhibits strong predictive performance while drastically reducing computational costs. We further investigate its robustness under changes in the distribution of economic scenarios by considering variations in the parameters of the underlying model of risk factors while the surrogate model is kept fixed. These results make the proposed approach particularly suitable for large-scale sensitivity analyses and fast balance-sheet evaluations in practical actuarial applications.
\newline
\newline
\setlength{\leftskip}{0cm} 
\setlength{\rightskip}{0cm} 

\section{Introduction}
\label{intro}

Since the enforcement of the Solvency II directive, European insurers are required to value their assets and liabilities on a market-consistent basis. While the market-consistent valuation of assets is generally straightforward, the absence of a liquid market for insurance liabilities makes their valuation much more challenging. Relying on pricing theory in fair and complete markets, the regulator defines the Best Estimate (BE) as the market-consistent value of liabilities which is computed as the expectation -- under a risk-neutral probability measure -- of the cashflows generated by the insurer’s portfolio of in-force policies. In the same spirit, the Value In Force (VIF) is defined as the expectation of the discounted future cashflows generated by the in-force portfolio for shareholders. In practice, at any given future time, the cashflow is not solely determined by the current values of the risk factors, but depends on their entire past trajectory (e.g. through the profit-sharing reserve). As a result of this path-dependent behavior, the expectation of each discounted cashflow cannot be computed analytically, and the computation of the expectation of the sum of discounted cashflows cannot be parallelized across time steps. Insurers therefore rely on Monte-Carlo methods to approximate it. The first step consists of using an Economic Scenario Generator (ESG) which generates a sample of economic scenarios, each of which consists of financial risk factors projected over time. Then each Monte-Carlo scenario of cashflows is obtained by running an asset-liability management (ALM) model on the corresponding Monte-Carlo economic scenario. ALM models involve complex management rules, profit-sharing mechanisms, surrender behaviors, and regulatory constraints, which introduce strong nonlinearities, barriers, and path-dependence in the cashflow generation process. The main computational challenge does not arise from a single model evaluation, but from the number of Monte-Carlo scenarios. This challenge is further worsened by the inherently sequential nature of the computations: cashflows are generated through iterative updates along each economic scenario, where each time step depends on past states, precluding parallelization across time. In addition, the presence of conditional mechanisms (e.g., management rules or policyholder behaviors) further limits vectorization and efficient implementation, ultimately leading to computational bottlenecks. Consequently, evaluating the VIF can lead to a significant computational burden. This issue becomes even more critical when insurers need to recompute the VIF under many different financial and non-financial assumptions. First, the insurer has to comply with several accounting frameworks (Solvency II and IFRS) and each one requires its dedicated ALM projections. For example, within the Solvency II framework, a European insurer relying on the Standard Formula needs to compute the Solvency Capital Requirement (SCR) and each SCR is obtained by aggregating the capital requirement associated with each risk module (market, life, etc.) and each risk sub-module (interest rates, mortality, lapses, etc.). This computational burden is even more pronounced for insurers using Internal Models. In this framework, the Solvency Capital Requirement is typically based on the distribution of the market-consistent balance sheet over a one-year horizon, which leads to nested valuation problems and requires repeated ALM projections under a large number of outer economic scenarios. This computational burden has motivated a substantial literature on acceleration methods for insurance liability valuation and solvency-capital calculation. Early contributions studied the direct nested-simulation problem and its acceleration through scenario selection and optimized allocation of inner simulations \citep{gordy2010nested,broadie2011efficient}. A second line of work replaces the full nested valuation by proxy models, including Least-Squares Monte-Carlo, curve fitting, and replicating portfolios \citep{bauer2010solvency,floryszczak2016inside,krah2018least,pelsser2016difference}. More recently, multilevel Monte-Carlo methods have also been proposed to reduce the cost of nested computations \citep{alfonsi2021multilevel,giles2019multilevel,boumezoued2025optimized}. In practice, insurers already use several levers to simplify and accelerate balance-sheet computations such as the number of scenarios, the number of model points -- i.e., the granularity at which liabilities (policyholder characteristics) and assets (individual asset positions) are represented -- and the time granularity. In this paper, we consider another type of approach which focuses on accelerating the computation of the sum of the discounted cashflows along a given economic scenario rather than focusing on ways to reduce the number of economic scenarios.
\newline

Let us now introduce a theoretical framework for the computation of the VIF when cashflows are generated by an ALM model. Although we focus on this specific indicator for clarity, the proposed framework extends straightforwardly to other ALM quantities of interest, such as the BE. Let $T>0$ be a finite time horizon and let $(\Omega, \mathcal{F}, \mathbb{P}^* ;\left(\mathcal{F}_t\right)_{t \leq T})$ be a filtered probability space, where $\mathbb{P}^*$ is a risk-neutral probability. We denote by $X=\left(X_t\right)_{t \in[0, T]}$ a continuous stochastic process taking values in $\mathbb{R}^d$ representing the financial risk factors driving the economic scenarios, and by $\left(\text{P\&L}_t\right)_{t \in[0, T]}$ the stochastic process describing the insurer's profit and loss. Both processes are assumed to be adapted to the filtration $(\mathcal{F}_t)_{t \leq T}$. We denote by $\mathcal{C}^0\left([0, T], \mathbb{R}^d\right)$ the set of continuous paths from $[0, T]$ to $\mathbb{R}^d$. In addition to financial risk factors, the valuation depends on two vectors $Z_{\rm f}$ and $Z_{\rm nf}$, gathering respectively the initial financial assumptions available at time $t=0$ (yield curves, implied volatilities, etc.) and non-financial assumptions (mortality rates, model points, surrender rates, etc.). We consider that the financial assumptions only affect the risk-neutral distribution of economic scenarios and the non-financial assumptions only affect the specification of the ALM model itself. For a given vector $Z_{\rm f}$, we denote by $\mathcal{L}(X(Z_{\rm f}))$ the corresponding risk-neutral distribution of sample paths of $X$. As we already mentioned, the ALM model maps the path of financial risk factors to the profit and loss generated by the insurer. Formally, for all $Z_{\rm nf}$, we assume the existence of a functional

\[
\varphi_{Z_{\rm nf}}:[0, T] \times \mathcal{C}^0\left([0, T], \mathbb{R}^d\right) \rightarrow \mathbb{R}
\]
such that

\[
\text{P\&L}_t=\varphi_{Z_{\rm nf}}\left(t, X_{[0, t]}\right),
\]
where $X_{[0, t]}: s \mapsto X_{s \wedge t}$ denotes the path of risk factors observed up to time $t$. For simplicity, we assume that the risk-free deflator is given by the first coordinate $\left(X_t^{\word{1}}\right)_{t \leq T}$. The sum of discounted cashflows generated by the ALM model is then defined as

\[
\text{CF}=\int_0^T X_t^{\word{1}} \text{P\&L}_t d t=\int_0^T X_t^{\word{1}} \varphi_{Z_{\rm nf}}\left(t, X_{[0, t]}\right) d t=\Phi_{Z_{\rm nf}}(X)
\]
where $\Phi_{Z_{\rm nf}}$ denotes the functional mapping a scenario $X$ to the corresponding integral of discounted cashflows. Such a functional representation can be found in \citep{cherchali2021modelisation}. In practice, cashflows are generated and observed on a discrete time grid rather than in continuous time so the integral becomes a sum. In the following, we will omit the dependence on $Z_{\rm nf}$ since we will focus on learning the mapping $\Phi_{Z_{\rm nf}}$ for a given non-financial assumption $Z_{\rm nf}$: hence we will denote $\Phi:=\Phi_{Z_{\rm nf}}$ and $\varphi:=\varphi_{Z_{\rm nf}}$. The VIF is therefore given by the risk-neutral expectation

\[
\text{VIF}=\mathbb{E}^*[\mathrm{CF}]=\mathbb{E}^*\left[\Phi(X)\right]
\]
with $X \sim \mathcal{L}(X(Z_{\rm f}))$ under $\mathbb{P}^*$. We recall that in a Monte-Carlo approach, the sample paths of $X$ are generated by an Economic Scenario Generator. Even if one assumes that the profit and loss process is Markovian with respect to the current risk factors, i.e. $\text{P\&L}_t=\varphi\left(t, X_t\right)$, the sum of discounted cashflows remains a non-trivial functional of the entire path of financial variables.
\newline

In this context, developing an accurate surrogate model for the functional $\Phi$, capable of efficiently approximating discounted cashflows along simulated economic scenarios, can be useful for several actuarial and risk management tasks. The purpose of such a surrogate is not to replace the full ALM model for regulatory balance sheet production, but rather to complement it in exploratory analyses, fast sensitivity computations, and model risk investigations. Typical use cases include the following:
\begin{enumerate}
\item The assessment of sensitivities with respect to financial assumptions, such as a recalibration of the ESG model. In these situations, the main bottleneck is not the generation of economic scenarios itself, but the repeated evaluation of the ALM model on each scenario.

\item The use of the surrogate in nested simulation settings, where ALM projections are required along many outer scenarios.

\item The generation of additional approximate cashflow evaluations at low cost, thereby increasing the effective sample size available for estimating quantities such as the VIF and reducing the corresponding Monte-Carlo variance.

\item The use of the surrogate as a sandbox model before launching full ALM production runs.
\end{enumerate}

An economic scenario containing the risk-free curve across 30 maturities and a basket of 5 equities, projected over a 50-year horizon with annual time steps, takes its values in $\mathbb{R}^{35 \times 50}\cong \mathbb{R}^{1750}$. This illustrates the high-dimensional nature of the input space on which the ALM cashflow functional is defined. Such high dimensionality raises two related difficulties for surrogate modeling. First, the approximation class must be expressive enough to capture the nonlinear and path-dependent response of the ALM model to the full economic scenario. Second, the estimation of this approximation must remain statistically and computationally feasible with a limited number of Monte-Carlo observations. These difficulties become even more pronounced when credit risk is included through market spreads and rating transition probabilities, since the number of risk factors may increase substantially. Although the ALM model is formally defined on this high-dimensional scenario space, its effective dimension is often much smaller. Many components of the economic scenario are strongly correlated, redundant, or only weakly relevant for the final discounted cashflow functional. Moreover, the raw scenario contains variables of heterogeneous nature and granularity: full term structures across maturities, equity trajectories, discount factors, credit spreads, risky curves and rating transition quantities. Treating all these coordinates as independent inputs would therefore lead to a high-dimensional and highly redundant learning problem. We emphasize that the high dimensionality of the scenarios is not, by itself, the main source of the computational cost of ALM projections: the bottleneck remains the repeated evaluation of the full ALM model across many scenarios. However, the dimension of the scenario space is a central issue for the construction of an accurate and statistically stable surrogate model. This motivates a preliminary preprocessing step before learning the surrogate. Its role is to construct a lower-dimensional version of the raw economic scenarios by exploiting prior knowledge of the structure of financial risk factors. For instance, yield curves are strongly correlated across maturities and are naturally summarized by a small number of curve factors, while some quantities generated by the ESG may be deterministic or nearly deterministic functions of others. In this paper, we separate the preprocessing of the economic scenario from the choice of the surrogate model. All benchmark models considered in the numerical section are trained on the same reduced representation of the economic scenarios. This ensures that the comparison focuses on the choice of the surrogate architecture rather than on differences in input engineering.
\newline
\newline
\textbf{Contributions.} In this work, we use a mathematical
object, called the time-augmented signature, which allows us to encode a continuous path in an efficient and parsimonious vector of features, with interesting approximation properties. More precisely, any continuous functional of a path can be approximated arbitrarily well by a finite linear combination of signature terms \citep{cuchiero2022universal, cuchiero2023global}. We leverage the approximation property of linear combinations of the truncated time-augmented signature to learn the sum of discounted cashflows as a deterministic response of the path of financial risk factors. In particular, we approximate $\Phi(X)$ by $\langle \beta, \widehat{\mathbb{X}}_{T} \rangle$, where $\beta$ is the vector of regression coefficients and $\widehat{\mathbb{X}}_{T}$ is the time-augmented signature of the (transformed) path of risk factors. More precisely, the signature is computed after a common preprocessing step that reduces the dimension of the raw economic scenarios, followed by optional path transformations designed to improve the expressiveness of the signature terms. The preprocessing is agnostic to the parametric specification of the ESG in the following sense: it only uses the simulated paths of economic quantities delivered by the ESG, together with the financial structure of these quantities, and does not rely on the stochastic equations or calibration procedure used to generate them. However, the learned reduction map is part of the fitted surrogate pipeline and is estimated from the training scenarios only. Our contribution is therefore not the mere use of dimension reduction, but the use of signature-based features, computed after this common preprocessing step, to approximate the ALM cashflow functional $\Phi$. We test this method on the reproducible ALM model specified in \citep{cherchali2021modelisation}.
\newline
\newline
\textbf{Related work.} The problem of predicting the output of a real-valued function whose input is a path has long been studied in the framework of functional linear regression \citep{ramsay1991some,marx1999generalized,ramsay1997functional,ferraty2006nonparametric,morris2015functional}. In this setting, the response is approximated by a linear functional of the input path: each coordinate of the path is weighted over time by an unknown coefficient function, which is itself represented in practice through a finite-dimensional basis expansion. While this approach provides an interpretable and well-established benchmark, it can be regarded as a first-order linear approximation of the functional. The signature framework offers a systematic way to go beyond this linear regime by constructing features that are universal for nonlinear functionals. There exists a substantial body of literature on the use of the signature of a path as an efficient family of features in machine learning \citep{levin2016learningpastpredictingstatistics, chevyrev2016primer, kiraly2019kernels, FERMANIAN2022105031} and quantitative finance \citep{arribas2020sig, cuchiero2023signature, bayer2023primal, bayer2025pricingamericanoptionsrough}, covering both theoretical foundations and practical applications. To our knowledge, the use of the signature as a family of features to approximate the discounted cashflows CF as a functional response of the path of financial risk factors has not yet been studied in the insurance industry. However, the signature has already been successfully used to address actuarial problems: we can cite \citep{andres2024signature} who studies an goodness-of-fit test based on the expected signature of two samples of stochastic processes for the validation of real-world economic scenarios, and \citep{yap2025forecasting} which leverages the universality properties of the signature to forecast mortality rate curves. In the actuarial literature, the vast majority of the papers focusing on the reduction of computation time of the balance sheet can be divided into regress-now and regress-later approaches \citep{glasserman2004simulation,wolf2025openirm}. A typical regress-now approach is the Least-Squares Monte-Carlo (LSMC) methodology, which learns the dependence of balance-sheet quantities such as the Best Estimate on the underlying financial and non-financial assumptions (represented by $Z_{\rm f}$ and by $Z_{\rm nf}$, defined above) from a collection of simulated scenarios \citep{lsmc}. On the other hand, regress-later approaches are used to regress the value of the cashflows on a set of basis functions. An example of a regress-later approach is the replicating portfolio method which uses cashflows of primary assets as a set of basis functions to regress the insurer's cashflows \citep{pelsser2016difference}. This method implicitly assumes that the insurer's cashflows can be statically hedged with a basket of non-path-dependent financial products, which is generally not the case \citep{albrecher2018asset}.
\newline
\newline
\textbf{Outline of the paper.} We proceed as follows. We briefly introduce the time-augmented signature of a path and we present our signature-based approximation in Section \ref{part1}. We discuss the ALM model and the ESG used in Section \ref{ALM model}. Numerical results are presented in Section \ref{part2}.

\section{Our approach to cashflow approximation}
\label{part1}

Taking the notations of Section \ref{intro}, we denote by CF the sum of discounted value of the profit and losses of the insurer, which admits the representation $\text{CF} = \Phi(X)$ with $\Phi : \mathcal{C}^{0}([0,T],\mathbb{R}^d) \rightarrow \mathbb{R}$ a path functional. The aim of this section is to introduce the signature of a path and show how it can be used to approximate the map $\Phi$.

\subsection{The signature of a path}

For a given path $X:=(X^{\word{1}},\cdots,X^{\word{d}}) :[0,T] \rightarrow \mathbb{R}^d$ in the set $BV([0,T],\mathbb{R}^d)$ of continuous paths with finite variation, and for each word $\word{i_1 \cdots i_n}$ with each letter $\word{i_k}$ taking its value in the alphabet $\mathcal{A}=\left \{\word{0},\word{1},\cdots,\word{d}\right \}$, we define

\[
\widehat{\mathbb{X}}^{\word{i_1 \cdots i_n}}_T := \int_{0}^{T} \int_{0}^{t_1} \cdots \int_{0}^{t_{n-1}} d\widehat{X}_{t_n}^{\word{i_1}} \cdots d\widehat{X}_{t_1}^{\word{i_n}},
\]
with $\widehat{X}_t^{\word{0}}=t$ and $\widehat{X}_t^{\word{i}}=X^{\word{i}}_t$ for $\word{i} \in \{ \word{1},\cdots,\word{d}\}$. Moreover, if we denote by $\emptyword$ the empty word which is the only element of $\mathcal{A}^{0}$, we define:

\[
\widehat{\mathbb{X}}^{\emptyword}_T:= 1.
\]
The time-augmented signature of $X$ is defined as $\widehat{\mathbb{X}}_T = \left(\widehat{\mathbb{X}}^{\word{i_1 \cdots i_n}}_T\right)_{(\word{i_1}, \cdots, \word{i_n}) \in \mathcal{A}^n, n \in \mathbb{N}}$ where $\mathbb{N}$ denotes the set of non-negative integers. For a given truncation order $N \in \mathbb{N}$, we can define the $N$-step truncated time-augmented signature of $X$ as $\widehat{\mathbb{X}}^{\le N}_T:=\left(\widehat{\mathbb{X}}^{\word{i_1 \cdots i_n}}_T\right)_{(\word{i_1}, \cdots, \word{i_n}) \in \mathcal{A}^n, n \le N} \in \mathbb{R}^{\frac{(d+1)^{N+1}-1}{d}}$.

\begin{example}
We consider a bounded variation path $(X_t)_{t \in [0,T]}: [0,T] \to \mathbb{R}$. The components of its $2$-step truncated time-augmented signature are 

\begin{align*}
 \begin{cases} 
\widehat{\mathbb{X}}_{T}^{\emptyword}=1 \\ 
\widehat{\mathbb{X}}_{T}^{\word{0}} = \int_{0}^{T} dt = T, \\
\widehat{\mathbb{X}}_{T}^{\word{1}} = \int_{0}^{T} dX_{t} = X_{T} - X_{0}, \\
\widehat{\mathbb{X}}_{T}^{\word{00}} = \int_{0}^{T} \int_{0}^{t} ds \, dt = \frac{T^2}{2}, \\ 
\widehat{\mathbb{X}}_{T}^{\word{01}} = \int_{0}^{T} \int_{0}^{t} ds  dX_t = \int_{0}^{T} t \, dX_t, \\
\widehat{\mathbb{X}}_{T}^{\word{10}} = \int_{0}^{T} \int_{0}^{t} dX_{s} dt = \int_{0}^{T} (X_{t} - X_0) dt, \\
\widehat{\mathbb{X}}_{T}^{\word{11}} = \int_{0}^{T} \int_{0}^{t} dX_s dX_t = \frac{(X_T - X_0)^2}{2}.
\end{cases}
\end{align*}  
\end{example}
The time-augmented signature of a path has a very powerful property: linear combinations of the components of the time-augmented signature are universal approximators. For clarity, we recall here the original formulation of the Universal Approximation Theorem in the framework of paths of bounded variation with the uniform topology which can be found in \citep{cuchiero2022universal}, even if other formulations exist when the compactness assumption is relaxed \citep{cuchiero2023global,bayer2023primal} and when the approximation is in an $L^p$ sense \citep{bayer2023primal,bayer2025pricingamericanoptionsrough} using weighted spaces and the notion of robust signature. We denote by $BV_0([0,T],\mathbb{R}^d)$ the set of paths of bounded variation starting from $0$ at $t=0$.

\begin{theorem}[Universal Approximation Theorem]
\label{UAT}
Let $\Psi : BV_0([0,T],\mathbb{R}^d) \rightarrow \mathbb{R}$ be a continuous functional for the uniform topology and $K \subset BV_0([0,T],\mathbb{R}^d)$ be a compact subset. Then for any $\varepsilon >0$, there exists $N_{\varepsilon} \in \mathbb{N}$, $\beta_{\varepsilon} \in \mathbb{R}^{\frac{(d+1)^{N_{\varepsilon}+1}-1}{d}}$ such that:

\begin{align*}
    \sup_{X \in K} \left| \Psi(X) - \left<\beta_{\varepsilon}, \widehat{\mathbb{X}}_{T}^{\le N_{\varepsilon}}\right>\right| \le \varepsilon.
\end{align*}
\end{theorem}
An important property of the time-augmented signature is its invariance with respect to translations, since it depends only on the increments of the path. As a consequence, the absolute level of the path is not captured by the time-augmented signature itself. To circumvent this issue, it is common in practical applications to add the origin $0$ at the beginning of the data stream \citep{chevyrev2016primer}. This transformation, known as the base-point augmentation removes, the translation invariance of the time-augmented signature transformation and embeds every path of bounded variation in  $BV_0([0,T],\mathbb{R}^d)$. Hence, throughout the remainder of this paper, the time-augmented signature will always be understood as being computed after applying the base-point augmentation. The Universal Approximation Theorem motivates the use of the time-augmented signature components as a family of features in a linear regression task. In this setting, one approximates the functional $\Psi$ by a linear model of the form
\[
X \mapsto \langle\beta, \widehat{\mathbb{X}}_{T}^{\le N}\rangle,
\]
where $\beta$ is a vector of coefficients to be learned from data. In practice, one does not fix a target precision $\varepsilon > 0$ and derive the corresponding truncation level $N_\varepsilon$. Instead, one selects a truncation level $N \in \mathbb{N}$ a priori and considers the best linear approximation of $\Psi$ within the finite-dimensional space spanned by truncated time-augmented signature components. More precisely, for a stochastic process $(X_t)_{t \in [0,T]}$, one aims to compute
\[
\beta^{*}_{N} := \text{argmin}_{\beta}  \mathbb{E}\left[\left|\Psi(X) - \beta^{\top} \widehat{\mathbb{X}}_{T}^{\le N}\right|^2\right].
\]
Since a linear combination of components of $\widehat{\mathbb{X}}_{T}^{\le N}$ is a particular case of a linear combination of components of $\widehat{\mathbb{X}}_{T}^{\le N+1}$, it implies that the corresponding best achievable approximation error is non-increasing as $N$ grows. This naturally motivates considering sufficiently large truncation levels. However, $\beta^{*}_{N}$ is unknown in practice, as the expectation cannot be computed exactly. Given a dataset $(X^{(k)}, Y^{(k)})_{1 \le k \le n}$, with $Y^{(k)} = \Psi(X^{(k)})$, one instead constructs an estimator $\widehat{\beta}_{N}$ by solving the empirical risk minimization problem (see below). In this context, taking $N$ arbitrarily large is not desirable: while increasing $N$ reduces the approximation error, it also increases the dimensionality of the feature space, which may deteriorate the statistical properties of the estimator. The choice of $N$ therefore results from a trade-off between approximation capacity and statistical estimation. We now turn to the empirical estimation of $\beta^{*}_{N}$. We use the following notations:

\begin{itemize}

\item $\mathbf{X} \in \mathbb{R}^{\frac{(d+1)^{N+1}-1}{d} \times n}$ denotes the design matrix such that for each $k \in \{1,\cdots,n\}$, the column $\mathbf{X}_{.,k}$ is equal to the components of the $N$-step truncated time-augmented signature of $X^{(k)}$,
\item $\mathbf{Y} \in \mathbb{R}^{1 \times n}$ denotes the response row vector such that for each $k \in \{1,\cdots,n\}$, the entry $\mathbf{Y}_k$ is equal to $\Psi(X^{(k)})$.

\end{itemize}
Finally, we can estimate $\beta^{*}_N$ by minimizing the following regularized empirical loss 

\begin{align}
\label{empirical loss}
\mathbb{R}^{\frac{(d+1)^{N+1}-1}{d}} \ni \beta \mapsto \| \mathbf{Y} - \beta^{\top}\mathbf{X}\|_{2}^{2} + \mathcal{R}(\beta),
\end{align}
where $\mathcal{R}$ is some regularization function. In practice, we can choose a hyperparameter $\lambda>0$ and define $\mathcal{R}(\beta)=\lambda \|\beta\|_{2}^2$ which leads to a Ridge regression, or $\mathcal{R}(\beta)=\lambda \|\beta\|_{1}$ which leads to a Lasso regression. The Ridge estimator admits a closed-form solution, while the Lasso typically requires iterative optimization procedures but can be desirable when seeking a sparse representation of the functional. 

A further important aspect is that the components of the time-augmented signature are linearly dependent \citep{bourdon2026linear}. In particular, for each truncation level $N$, the family of components indexed by words that do not start with the letter $\word{0}$ (referred to as suffixes) spans the same linear space as the full set of the $N$-step truncated time-augmented signature components. Consequently, for linear regression purposes, it is sufficient to restrict to this subset of features. This has several advantages:

\begin{enumerate}
     
    \item the storage cost of features is reduced: the cardinality of the subset of components goes from $\frac{(d+1)^{N+1}-1}{d}$ to $(d+1)^N$,
    \item the computational cost is reduced,
    \item the prediction performance is preserved in theory, while empirical improvements may arise in practice due to improved conditioning, reduced variance, or more stable cross-validation.
\end{enumerate}

In the remainder of the paper, we restrict ourselves to this subset of components and, by a slight abuse of notation, still denote it by $\widehat{\mathbb{X}}_{T}^{\le N}$. It turns out that this subset of features is linearly independent for the almost-sure equality when $(X_t)_{t \in [0,T]}$ is the affine interpolation of the solution to a SDE with additive Brownian noise. In practice, we observe that these features are linearly independent for more general stochastic processes. A consequence of the linear independence of this subfamily of components is the uniqueness of $\beta^{*}_{N}$. Hence, the optimal model is identifiable and this improves interpretability. Moreover, compared to other families of components with the same spanning property, the family of suffixes leads to the minimal computation time when the terms of the time-augmented signature are computed using Chen's relation.

\subsection{Implementation}
\label{implementation}

In this section, we present the detailed pipeline from the raw economic scenario to the final output. We emphasize that the raw economic scenario is both the output of the ESG and the input of the ALM model. 

\subsubsection{Dimension reduction as a common preprocessing layer}

In a realistic setting, the raw economic scenarios produced by the ESG contain many risk factors observed over a projection horizon. A naive construction of a surrogate model would consist of feeding this raw multivariate time series directly into a regression architecture. Such an approach is generally inappropriate in the insurance setting, for both statistical and computational reasons. The number of Monte-Carlo paths available for training is typically limited, whereas the raw economic scenarios contain many variables (typically several thousand if we multiply the number of risk factors by the number of time steps). Consequently, the effective sample size is small relative to the apparent input dimension. For signature-based models, this issue can be quantified explicitly. The number of features -- and hence the number of parameters to be estimated -- grows exponentially with the truncation order of the signature and polynomially with the dimension of the underlying path. In our numerical applications in Section~\ref{part2}, the raw economic scenarios contain 386 risk factors projected over a 50-year horizon. The suffix components of the 2-step truncated time-augmented signature therefore form a vector of size $(386+1)^2 = 149,769$ which already far exceeds the number of paths available in a standard insurance dataset, typically ranging from 1000 to 5000 observations; see \citep{arrouy2022economic} for practical considerations on ESG datasets. A naive linear regression, even with regularization, would therefore operate in a regime where the number of features is several orders of magnitude larger than the number of observations, leading to severe statistical and computational difficulties \citep{hastie2009elements,buhlmann2011statistics}. This feature explosion should not be interpreted as a difficulty specific to signatures. Other surrogate models also suffer from the direct use of high-dimensional economic scenarios. We therefore introduce a preliminary dimension-reduction map $f_{\theta}$ which transforms a raw economic scenario into a lower-dimensional path that preserves the main economic information. This map is used as a common preprocessing layer before applying any surrogate model. In particular, the benchmark models in Section~\ref{part2} are trained on the same reduced paths, so that differences in performance can be entirely attributed to the surrogate architecture rather than to different input representations.

The purpose of the map $f_\theta$ is to transform the raw economic scenarios -- which are the inputs of the ALM model -- into lower-dimensional paths before any surrogate model is fitted. This step is motivated by the financial structure of the economic scenarios. In particular, an economic scenario contains several blocks of variables whose dimensions are high because of the fine granularity of assets in ALM models, not because they represent independent sources of risk. For example, a yield curve scenario contains many maturities allowing a precise pricing of the bonds within the insurer asset portfolio, but the yield curve movements are typically driven by a few dominant factors \citep{litterman1991common}. Moreover, some risk factors are derived from others: for instance, discount factors can be reconstructed from the risk-free curve, and risky curves can be represented through the joint information contained in the risk-free curve and credit spreads. These redundancies in the scenario set can be thought of as a form of preprocessing since they avoid additional computations in the ALM model. In our numerical implementation, we therefore apply PCA to the different rate curves. The PCA coefficients define the parameter $\theta$ of the reduction map $f_\theta$ and are estimated on the training dataset only. We retain a single principal axis for each curve (the risk-free rate curve and a rate curve for each rating), corresponding to a level-type factor, although the framework allows one to retain several principal axes if needed. The principal axes and the associated level components are estimated by stacking all scenarios and all time steps. This choice reflects the empirical structure of the economic scenarios and does not require access to the equations of the ESG model. Hence, the preprocessing can be applied to any ESG producing the same type of economic quantities, regardless of its parametric specification. The economic scenarios considered in this paper incorporate credit risk through risky
rate curves and rating transition probabilities; see Section~\ref{part ESG}. We choose
not to keep the full set of rating transition probabilities as separate input
coordinates, since the most relevant information for the ALM model, namely the compensation for default risk, is already embedded in the credit spreads inferred from the risky curves. This is a modeling choice based on the economic interpretation of the economic scenarios.

The preprocessing is model-agnostic with respect to the ESG specification: it only requires economic scenarios produced by an ESG and does not use the stochastic dynamics that generated them. However, as with any data-driven preprocessing step, the numerical values of the reduction parameters $\theta$ are estimated from the training sample and are therefore part of the fitted surrogate pipeline. More details on the input risk factors can be found in Section \ref{part ESG}. After this dimension reduction step, the resulting paths contain 10 risk factors over 50 years: the risk-free discount factor, the level factor of the risk-free rate curve, the equity coordinate, and the level factor of each of the seven risky rate curves.

\subsubsection{Transformations}

In order to enhance the expressiveness of the features and improve statistical
learning, it is common to apply transformations to the path prior to computing its
truncated time-augmented signature
\citep{levin2016learningpastpredictingstatistics,chevyrev2016primer,kiraly2019kernels,FERMANIAN2022105031,andres2024signature}. Such transformations may lead to more effective representations of the path. A first
transformation is dictated by the financial nature of the underlying risk factor:
when considering an equity index, we systematically replace its level by its
log-returns, since the index follows a multiplicative dynamics and its returns are
the economically meaningful quantity, whereas its level is not directly comparable
across scenarios nor across time. Unlike the transformations described below, this
one is applied in every configuration considered in this paper. The remaining
transformations are optional, in the sense that their activation is treated as a
hyper-parameter of the surrogate model whose impact is studied numerically in
Section~\ref{part2}. The most commonly used ones include the lead-lag transform and
the cumulative transform. Note that the cumulative transform does not increase the
dimension of the path, whereas the lead–lag transform doubles the dimension of the
underlying path. More details on these transformations can be found in
\ref{appendix A}. We denote by $\varphi$ the deterministic map which takes the entire
(reduced) economic scenario and returns the transformed (reduced) economic scenario.

\subsubsection{Regression on the signature}

We use the Python library \texttt{iisignature} \citep{iisignature}, for which we implemented the efficient computation of the suffix family proposed in \citep{bourdon2026linear}. Once the suffix components of the truncated time-augmented signature of the transformed paths are computed, it is very standard to learn the expectation and the variance of these features before centering and scaling them. Indeed, this standardization technique is just an affine transformation of the features and then any linear combination of them can still be represented as an affine combination of the newly constructed features and reciprocally (the constant $1$ is part of the suffix components of the time-augmented signature and acts as an intercept in the regression). However, this transformation is of key importance for Ridge and Lasso regressions where the amplitude of $\beta$ in each direction has a direct impact on the regularized training loss. If two centered features do not have the same amplitude then their corresponding coefficients will not be penalized the same way: scaling is a method to prevent this issue. In an idealized setting, we would want access to the vector $\mu = (\mathbb{E}[ \widehat{\mathbb{X}}_{T}^{\word{w}}])_{\word{w} \in \mathcal{A}^k;1 \le k\le N}$ and the matrix $D = \text{diag}(\text{Var}(\widehat{\mathbb{X}}_{T}^{\word{w}}) ~|~ \word{w} \in \mathcal{A}^k; 1 \le k \le N)$ to construct the features $\mathcal{X}:=D^{-\frac{1}{2}}(\widehat{\mathbb{X}}_{T}^{\le N} - \mu)$. Note that the component $\widehat{\mathbb{X}}_{T}^{\emptyword}=1$ is not impacted since it plays the role of the intercept. The matrix $D$ and the vector $\mu$ are parameters of the model and are estimated with their empirical estimators on the training dataset during the training phase.

\subsubsection{Pipeline summary}

Finally, the full model pipeline consists of: 

\begin{enumerate}
    \item reducing the dimension of the underlying raw path of risk factors, 
    \item applying transformations to the reduced path,
    \item computing the suffixes components of the truncated time-augmented signature,
    \item centering and scaling these features to achieve zero mean and a unit variance,
    \item regressing the output on the features.
\end{enumerate}

\begin{figure}[H]
\hspace*{-1cm}
\begin{tikzpicture}[shorten >=1pt,draw=black!50, node distance=\layersep]

\draw[rounded corners=\corners, black, fill=green!30, thick] (0.3,0) rectangle ++(1.4,1); 
\node[text width=10em, text centered] at (1,0.5) {$(X_t)_{t \le T}$};
\node[text width=4em, text centered] at (1,1.5) {Reduced path}; \draw[black, thick, ->] (1.7,0.5) -- (2.5,0.5); 

\draw[rounded corners=\corners, black, fill=blue!30, thick] (2.5,0) rectangle ++(1.4,1); \node[text width=6em, text centered] at (3.2,1.5) {Transfor\-mations}; \node[text width=4em, text centered] at (3.2,0.5) {$\varphi$}; \draw[black, thick, ->] (3.9,0.5) -- (4.7,0.5);


\draw[rounded corners=\corners, black, fill=blue!30, thick] (4.7,0) rectangle ++(1.4,1);
\node[text width=6em, text centered] at (5.4,1.5) {Suffix\\components};
\node[text width=4em, text centered] at (5.4,0.5) {$\widehat{\mathbb{X}}_{T}^{\le N}$}; \draw[black, thick, ->] (6.1,0.5) -- (6.9,0.5); 


\draw[rounded corners=\corners, black, fill=blue!30, thick] (6.9,0) rectangle ++(3.5,1); \node[text width=10em, text centered] at (8.6,1.5) {Standardization}; \node[text width=10em, text centered] at (8.6,0.5) {$\mathcal{X} = D^{-\frac{1}{2}}(\widehat{\mathbb{X}}_{T}^{\le N} - \mu)$}; \draw[black, thick, ->] (10.4,0.5) -- (11.2,0.5);


\draw[rounded corners=\corners, black, fill=blue!30, thick] (11.2,0) rectangle ++(1.9,1); \node[text width=6em, text centered] at (12.15,1.5) {Affine combination}; \node[text width=10em, text centered] at (12.15,0.5) {$\beta_0 +\left<\beta,\mathcal{X}\right>$}; \draw[black, thick, ->] (13.1,0.5) -- (14,0.5); 


\draw[rounded corners=\corners, black, fill=red!30, thick] (14,0) rectangle ++(1.0,1); \node[text width=6em, text centered] at (14.5,1.5) {Predicted output}; \node[text width=4em, text centered] at (14.5,0.5) {$\widehat{\text{CF}}$}; 


\draw[dashed, thick, red, rounded corners=5pt] (6.7,-0.3) rectangle (13.4,2) node[above, midway, yshift=3.5em, text=red] {}; 


\draw[dashed, thick, rounded corners=5pt] (2.3,-0.3) rectangle (6.5,2) node[above, midway, yshift=3.5em,text=gray]{}; 
\end{tikzpicture} 
\centering
\caption{\textbf{Signature-based model: $G_{\text{sig}}$.} Parameters of the model: $\mu, D,\beta_0, \beta$. Hyper-parameters of the model: $\varphi,N$. The blocks circled in red correspond to the components of the model that are optimized during training, whereas the blocks circled in gray correspond to components specified prior to the training phase.}
\label{fig:signature_model} 
\end{figure}
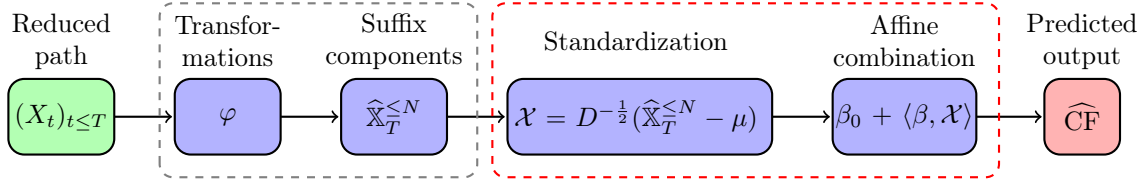

We distinguish the complete surrogate, which acts on the original economic scenario $X$, from the signature-specific regression map, which acts on the preprocessed path. We denote by $G_{\mathrm{sig}} : \mathcal C^0([0,T],\mathbb R^{d_\theta}) \longrightarrow \mathbb R$ the signature-based regression map applied after preprocessing, where $d_\theta$ is the dimension of the path obtained after applying $f_\theta$. It is defined by $G_{\mathrm{sig}}(X)=\beta_0 + \left\langle \beta, D^{-\frac{1}{2}}\left(S_N \circ \varphi(X) - \mu \right) \right\rangle$ for $X \in \mathcal C^0([0,T],\mathbb R^{d_\theta})$, where $S_N : X \mapsto \widehat{\mathbb{X}}_{T}^{\le N}$ denotes the $N$-step truncated time-augmented signature transform. The complete surrogate model applied to the original economic scenario is then $\Phi_{\mathrm{sig}}(X)=G_{\mathrm{sig}}\left(f_\theta(X)\right)$. We provide in Figure~\ref{fig:signature_model} a synthetic description of the signature-specific component $G_{\mathrm{sig}}$, while the common preprocessing map $f_\theta$ is deliberately kept outside the diagram.

\paragraph{Training procedure.}

The truncation order $N$ and the transformation map $\varphi$ are not selected using cross-validation. In the following, we instead explore several values of $N$ and different configurations of $\varphi$, where $\varphi$ simply encodes whether the lead-lag and/or cumulative transformations are simultaneously applied to each risk factor. We then study their impact on both generalization performance and computational cost. The parameters of the model are estimated using the training dataset. In particular, the PCA projection coefficients $\theta$, as well as the standardization parameters $\mu$ and $D$, are estimated on the training dataset and kept fixed afterwards in order to avoid information leakage. The regression coefficients $\beta_0,\beta$ are obtained by minimizing the regularized empirical loss introduced in \eqref{empirical loss}. We choose two regularization strategies: a Ridge regression ($\mathcal{R}(\beta)=\lambda \|\beta\|_2^2$) and a Lasso regression ($\mathcal{R}(\beta)=\lambda \|\beta\|_1$). The regularization hyper-parameter $\lambda$ is selected by cross-validation on the training dataset.

\section{Description of the ALM model and the input risk factors}
\label{ALM model}

\subsection{Description of the ALM model}
We provide a concise overview of the ALM model specified in \citep{cherchali2021modelisation}, to which we have added a credit layer and made slight modifications. We emphasize that the model specifications are designed to be consistent with French regulatory requirements and French GAAP. The model tracks both the market and the book value of the assets in order to apply the regulatory profit-sharing mechanism. It also features a determination of the crediting rate that reflects market practice, resulting from a trade-off between a competitive rate, a regulatory rate, and available profits. The major extension consists of the integration of corporate bond investments with explicit credit risk modeling via a Jarrow-Lando-Turnbull (JLT) framework combined with rating transition matrices. The model considers an insurance company that has sold life insurance contracts to many policyholders. We assume that the contracts are mainly described by two drivers: the minimum guaranteed rate $r^G$ that triggers the minimal earnings, and the participation rate $\pi_{pr} \in [0, 1]$ that forces the insurer to redistribute this proportion of gains on equity assets. Policyholders do not receive intermediate payments as they are paid only when they exit the life insurance contract. Asset allocation strategy and the profit-sharing mechanism are two key components of an ALM model that are typically at the discretion of the insurer. As such, they constitute the main sources of variation across different ALM specifications and play a central role in the dynamics of the model. Before introducing the model dynamics, we specify the time framework. In practice, ALM projections are computed on a discrete time grid
\[
0 = t_0 < t_1 < \dots < t_K = T,
\]
where $t_{k+1} - t_k = 1$ year for all $k = 0, \dots, K-1$. To maintain consistency with the continuous-time framework introduced in Section \ref{intro}, all ALM quantities are modeled as continuous-time processes that are constant over each interval $[t_k, t_{k+1})$, while the underlying economic scenarios are linearly interpolated between grid points to ensure the well-posedness of their time-augmented signatures. We emphasize that the computational burden of this ALM model is not primarily driven by a single particularly expensive operation, but by the repeated sequential updates required along each economic scenario. In practice, the cost scales with the number of scenarios, the number of time steps, and the number of model points used to represent the insurer's portfolio. This is precisely the type of computational structure for which a pathwise surrogate model can be useful. We now describe the asset allocation strategy and the profit-sharing mechanism used in the model.

\subsubsection*{Asset allocation}

The initial deposit of the policyholder is invested in three asset classes according to a fixed portfolio allocation that remains constant over time:

\begin{enumerate}
\item[(i)] \textbf{Equities}: the allocation is denoted by $w_e$.

\item[(ii)] \textbf{Risk-free bonds}: the total allocation is denoted by $w_b$. At each rebalancing date, this allocation is distributed across a basket of zero-coupon bonds with maturities $1,2,\cdots,n_{\text{basket}}$ years. The weights assigned to each maturity are chosen so that the aggregate exposure matches the target allocation $w_b$.

\item[(iii)] \textbf{Corporate bonds}: for each rating $R \in \mathcal{R}$, where $\mathcal{R}$ denotes the set of considered ratings, we denote by $w_{b,R}$ the allocation to bonds with rating $R$. Similarly to the risk-free case, this allocation is distributed at each date across a basket of bonds with maturities $1,2,\cdots,n_{\text{basket}}$ years, so that the total exposure to rating $R$ is equal to $w_{b,R}$.
\end{enumerate}

\subsubsection*{Profit-sharing mechanism}

The profit-sharing reserve (PSR) accumulates realized gains in order to redistribute them to policyholders through the credited rate. For each $k=1,\cdots,K$, its value at time $t_k$ is recursively defined as:
\[
PSR_{t_k} = r_{t_k}^{\text{cred}} \cdot PSR_{t_{k-1}} + (1 - \rho^*_{t_k}) \cdot \left( PSR_{t_{k-1}} + \max\left\{ LGL^{\text{equity}}_{t_k}(\alpha_{t_k}) + RGL^{\text{equity}}_{t_k}, 0 \right \} \right)
\]

with initial condition $PSR_{0} = 0$, where:
\begin{enumerate}
    \item $r^{\text{cred}}_{t_k}$ denotes the credited rate applied to the previous value of the profit-sharing reserve,
    \item $\rho^*_{t_k} \in [0,1]$ denotes the share of the profit-sharing reserve that is released,
    \item $LGL^{\text{equity}}_{t_k}(\alpha_{t_k}):=\alpha_{t_k}\max\left\{LGL^{\text{equity}}_{t_k},0\right\} - (1-\alpha_{t_k})\max\left\{-LGL^{\text{equity}}_{t_k},0\right\}$ denotes the partial recognition of the latent gains/losses on equities $LGL^{\text{equity}}_{t_k}$, parametrized by $\alpha_{t_k} \in [0,1]$ defined in \eqref{eq alpha} below,
    \item $RGL^{\text{equity}}_{t_k}$ denotes the realized gains/losses on equities.
\end{enumerate}
The credited rate distributed to policyholders takes into account:
\begin{enumerate}
    \item the guaranteed minimum rate $r^G$,
    \item the competitive rate $r^{\text{comp}}_{t_k}$ which is the estimation of the rate credited by competitors,
    \item smoothing through the profit-sharing reserve in order to mitigate year-to-year fluctuations.
\end{enumerate}
For $k=1,\cdots,K$, we denote by $(\alpha,\rho) \mapsto TD_{t_k}(\alpha, \rho)$ the distributable surplus function at time $t_k$, defined as follows:
\[
TD_{t_k}(\alpha, \rho) = LGL^{\text{equity}}_{t_k}(\alpha) + BV^{\text{equity}}_{t_k} \cdot \left(1 + \pi_{pr} \cdot \rho \right),
\]
where $BV_{t_k}^{\text{equity}}$ denotes the book values of equities at time $t_k$. The credited rate is determined according to four cases, depending on performance thresholds:
\[
r_{t_k}^{\text{cred}} =
\begin{cases} 
\frac{\pi_{pr} \cdot TD_{t_k}(0, \rho^*_{t_k})}{MR_{t_k} + PSR_{t_{k-1}}} 
&  \text{if Case A: } \pi_{pr} \cdot TD_{t_k}(0, \rho^*_{t_k}) \geq \max(r^G, r^{\text{comp}}_{t_k}), \\[0.5em]

\frac{\pi_{pr} \cdot TD_{t_k}(\alpha_{t_k}, \rho^*_{t_k})}{MR_{t_k} + PSR_{t_{k-1}}} 
& \text{if Case B: } \pi_{pr} \cdot TD_{t_k}(0, \rho^*_{t_k}) < \max(r^G, r^{\text{comp}}_{t_k}) \leq \pi_{pr} \cdot TD_{t_k}(1, \rho^*_{t_k}), \\[0.5em]

\frac{\pi_{pr} \cdot TD_{t_k}(1, \rho^*_{t_k})}{MR_{t_k} + PSR_{t_{k-1}}} 
& \text{if Case C: } r^G \leq \pi_{pr} \cdot TD_{t_k}(1, \rho^*_{t_k}) < \max(r^G, r^{\text{comp}}_{t_k}), \\[0.5em]

\max\left\{ r^G, \frac{\pi_{pr} \cdot TD_{t_k}(1,1)}{MR_{t_k} + PSR_{t_{k-1}}} \right\}
& \text{if Case D: } \pi_{pr} \cdot TD_{t_k}(1, \rho^*_{t_k}) < r^G,
\end{cases}
\]
where $MR_{t_k}$ denotes the mathematical reserves and $\alpha_{t_k}$ is defined as
\begin{align}
\label{eq alpha}
\alpha_{t_k}:=\begin{cases}
0  &\text{if Case A}, \\
\frac{\max \{r^G,r^{\text{comp}}_{t_k}\} - \pi_{pr} \cdot TD_{t_k}(0, \rho^*_{t_k})}{\pi_{pr}[TD_{t_k}(1, \rho^*_{t_k}) -  TD_{t_k}(0, \rho^*_{t_k})]} &\text{if Case B}, \\
1 &\text{if Case C},\\
1 &\text{if Case D}.
\end{cases}
\end{align}

The different cases reflect the regulatory and economic constraints faced by the insurer when setting the credited rate:

\begin{enumerate}
    \item Case A corresponds to a situation where the available distributable surplus is sufficiently high to meet both the guaranteed rate and the competitive rate without mobilizing latent gains. In this case, the insurer can afford to smooth returns by retaining part of the surplus within the profit-sharing reserve (i.e., $\alpha_{t_k} = 0$).
    \item Case B represents an intermediate situation where the surplus is not sufficient in its base form but becomes adequate when partially recognizing latent gains on equities. The insurer therefore adjusts the parameter $\alpha_{t_k} \in (0,1)$ to meet the target rate $\max(r^G, r^{\text{comp}}_{t_k})$ while limiting the realization of latent gains.
    \item Case C arises when even the full recognition of latent gains is insufficient to reach the competitive rate, but still allows the insurer to meet the guaranteed rate. In this case, all available gains are mobilized ($\alpha_{t_k} = 1$), and the credited rate remains below the market benchmark but above the guarantee.
    \item Case D corresponds to a stressed situation where even the full mobilization of available surplus is insufficient to meet the guaranteed rate. The insurer is then constrained to credit at least the guaranteed rate $r^G$, by drawing on the profit-sharing reserve.
\end{enumerate}
Once the asset allocation and the profit-sharing mechanism have been specified, we can describe how the profit and loss cashflows are constructed. For each $k = 1,\dots,K$ we define the profit and loss evaluated at $t_k$ as follows:

\begin{align}
\label{p&l}
\text{P\&L}_{t_k}
&= AM_{t_k} + CR_{t_{k-1}}\left(\frac{1}{P_{t_{k-1}}(t_k)} - 1\right)
+ CR_T \mathbbm{1}_{\{t_k = T\}}.
\end{align}
The profits and losses represent the net accounting result over a period and combine:

\begin{enumerate}
\item the accounting margin $AM_{t_k}$, i.e. the difference between the book value of assets
(equities, risk-free and corporate bonds, cash, etc.) and liabilities
(mathematical reserves and profit-sharing reserve) before reserve
adjustments,
\item the effect of the previous capitalization reserve $CR_{t_{k-1}}$
reinvested at the short rate given by the one-year risk-free zero-coupon bond $P_{t_{k-1}}(t_k)$,
\item the capitalization reserve $CR_T$ obtained at the end of the projection after full liquidation of the portfolio.
\end{enumerate}
The process $(P_{t_{k-1}}(t_k))_{k=1,\cdots,K}$ is given by the economic scenarios, while the processes $(AM_{t_k})_{k=1,\cdots,K}$ and $(CR_{t_k})_{k=0,\cdots,K}$ are constructed by the ALM model.

\subsubsection*{Dynamics of the capitalization reserve}

The process $(CR_{t_k})_{k=0,\cdots,K}$ satisfies $CR_{0}=0$ and for all $k=1,\cdots,K$:

\[
CR_{t_k} = \max\left\{ CR_{t_{k-1}} + RGL_{t_k},\;0\right\},
\]
where $RGL_{t_k}$ denotes the total realized gains/losses over all assets.

\subsubsection*{Dynamics of the accounting margin}

The accounting margin $(AM_{t_k})_{k=1,\dots,K}$ is defined for all $k=1,\dots,K$ by
\[
AM_{t_k} = (1-\pi_{pr})TD_{t_k}(\alpha_{t_k},\rho^{*}_{t_k}) - \max\{R_{t_k}^{G}(\alpha_{t_k},\rho^{*}_{t_k})-\pi_{pr}TD_{t_k}(\alpha_{t_k},\rho^{*}_{t_k}),0\},
\]
with
\[
R_{t_k}^{G}(\alpha,\rho):=\max \{ r^{G}(MR_{t_k}+PSR_{t_k -1}), \pi_{pr} TD_{t_k}(\alpha,\rho)\}.
\]
Finally, the variable CF is defined as:

\begin{align}
\label{CF alm}
\text{CF}=\sum_{k=1}^{K} \left(\prod_{m=0}^{k-1}P_{t_m(t_{m+1})}\right) \text{P\&L}_{t_k},
\end{align}
with $(\text{P\&L}_{t_k})_{k=1,\cdots,K}$ defined in \eqref{p&l}. Parameters used for the specification of the ALM model are reported in \ref{appendix B}.

\subsection{Description of the Economic Scenario Generator}
\label{part ESG}

Consistently with the asset allocation strategy, we consider raw economic scenarios that consist of the following risk factors:

\begin{enumerate}
    \item the risk-free discount factor,
    \item the risk-free rate curve -- with $40$ maturities,
    \item an equity index,
    \item risky rate curves -- each one with $40$ maturities,
    \item probabilities of rating transition.
\end{enumerate}
It follows that a raw economic scenario contains 386 risk factors. These risk factors are projected over a 50-years time horizon with annual time steps. In order to generate a whole economic scenario, it is in fact sufficient to explicitly model only the risk-free rate curve, the equity index, and the rating transition probabilities, as the remaining quantities can be derived from them. More precisely, the risk-free discount factor is obtained as the value process of a portfolio that is rolled over annually in one-year zero-coupon bonds, and can therefore be reconstructed from the risk-free curve. Similarly, each risky rate curve can be expressed as the sum of the risk-free rate curve and a credit spread. This spread reflects the compensation for default risk and is typically modeled as a function of the default probability, so that risky curves can be inferred from the joint dynamics of the risk-free curve and the probabilities of rating transition. We provide in Table \ref{tab:risk factors models} the models used for each risk factor.

\begin{table}[H]
    \centering
    \begin{tabular}{l|l}
         Risk factor & Model \\
         \midrule
         Risk-free rate curve ($\Rightarrow$ risk-free discount factor) & Displaced Diffusion Libor Market Model \\
         Equity & Black-Scholes with time-dependent volatility\\
         Probabilities of rating transition ($\Rightarrow$ spreads $\Rightarrow$ risky rate curves) & Jarrow-Lando-Turnbull
    \end{tabular}
    \caption{Risk factor models.}
    \label{tab:risk factors models}
\end{table}

\subsubsection*{Risk-Free Rate Model}

The risk-free rate curve is obtained using a Displaced Diffusion Libor Market Model (DDLMM), which directly specifies the dynamics of forward rates. The spot rate curve can then be uniquely recovered from the forward curve via no-arbitrage relationships. For each $k=1,\cdots,K$, we denote by $F_t^{(k)}$ the forward rate for the period $[t_k, t_{k+1}]$ at time $t < t_k$. We set $F_t^{(k)}=\tilde{F}_t^{(k)}- \delta$ with $\tilde{F}_t^{(k)}$ the displaced forward rate and $\delta>0$ allowing for negative forward rates. Under the $k+1$-forward neutral probability associated with the numeraire $(P_t(t_{k+1}))_{t \in [0,t_k]}$, the dynamics of the displaced forward rate is the following:

\[
d\tilde{F}_t^{(k)} = \tilde{F}_t^{(k)} \sum_{j=1}^{M} \xi_{t}^{(k),j} dW_t^{(k+1),j},
\]
where $W^{(k+1)}:=(W^{(k+1),j})_{j=1,\cdots,M}$ is a $M$-dimensional standard Brownian motion under the $k+1$-forward neutral probability and $\xi^{(k)}:=(\xi^{(k),j})_{j=1,\cdots,M}$ is a deterministic function defined on $[0,t_k]$ taking values in $\mathbb{R}_+^{M}$. We assume that for all $k=1,\cdots,K$ and all $j=1,\cdots, M$, the following decomposition holds:
\[
\xi_{t}^{(k),j} = \sigma^{(k)}_t \beta^{(k),j}_t, \quad \forall t \in [0,t_k].
\]
The family of functions $(\beta^{(k),j})_{k=1,\cdots,K; j=1,\cdots,M}$ satisfies $\sum_{j=1}^{M} (\beta^{(k),j}_t)^2=1$ for all $t \in [0,t_k]$, for all $k=1,\cdots,K$. In order to reduce the number of degrees of freedom, these functions are assumed to be piecewise constant, depending on the time to maturity. The volatility structure $\sigma^{(k)}$ is decomposed into two parametric functions; see \citep{brigo2006interest} for more details. For all $k=1,\cdots,K$, we assume that:

\[
\sigma^{(k)}_t=H(t)g(t_{k+1} - t), \quad \forall t \in [0,t_{k}],
\]
with $H$ a time-dependent scaling factor which does not depend on $k$, defined as 
  
\[
H(t):=\theta + (1-\theta)e^{-\kappa t}, \quad \forall t \in [0,T],
\]
and $g$ is defined as

\[
g(u):= (bu + a)e^{-cu} + d, \quad \forall u \in [0,T].
\]
Moreover $(\kappa,\theta,a,b,c,d) \in \mathbb{R}_{+}^6$.

\subsubsection*{Equity Model}

The equity index $(S_t)_{t \in [0,T]}$ follows a Black-Scholes dynamics with time-dependent volatility:
\[
dS_t = r_t S_t dt + \sigma(t) S_t dW_t,
\]
where $\sigma : [0,T] \rightarrow \mathbb{R}^+$ is a deterministic function of time and $(W_t)_{t \in [0,T]}$ is a one-dimensional Brownian motion under $\mathbb{P}^*$. In particular, we assume that $\sigma$ is piecewise constant on each interval $[t_k,t_{k+1})$:

\[
\sigma := \sum_{k=0}^{K-1} \sigma^{(k)} \mathbbm{1}_{[t_k,t_{k+1})},
\]
with $(\sigma^{(k)})_{k=0,\cdots,K-1} \in \mathbb{R}_{+}^{K}$. The drift $r_t$ is consistent with the risk-free rate.

\subsubsection*{Credit Risk Model}

Credit risk is modeled using the Jarrow-Lando-Turnbull (JLT) framework. We consider the finite set of ratings $\mathcal{R}$ where $R_d \in \mathcal{R}$ corresponds to the default state. The model relies on a reference (time-homogeneous) annual transition probability matrix $Q = (Q^{r,r'})_{(r,r') \in \mathcal{R}^2}$: its entry $Q^{r,r'}$ denotes the probability that a transition occurs from state $r$ to state $r'$ over one year. The related generator matrix $\Lambda$ is defined as 

\[
\Lambda:=\log(Q) =\sum_{n=1}^{\infty} (-1)^{n+1} \frac{(Q-I_{\text{Card}(\mathcal{R})})^{n}}{n}
\]
We assume that $Q$ is diagonalizable into $Q=MDM^{-1}$ with $D=\text{diag}(d_R ~|~ R \in \mathcal{R})$ and $d_R \in \mathbb{R}$ for each rating $R \in \mathcal{R}$. It follows that 

\[
\Lambda=M \log(D) M^{-1},
\]
with $\log(D)=\text{diag}(\log(d_R) ~|~ R \in \mathcal{R})$. The JLT model captures the dynamics of the transition probabilities among different rating groups, as well as transitions from any group to the default state $R_{d}$, by providing a stochastic adjustment to the generator $\Lambda$, transformed into:

\[
\Lambda_t = \pi_t \Lambda,
\]
where the so-called "risk-premium adjustment" $(\pi_t)_{t \in [0,T]}$ follows the dynamics:

\[
d\pi_t = \alpha(\mu - \pi_t)dt + \gamma\sqrt{\pi_t} dB_t,
\]
where $(B_t)_{t \in [0,T]}$ is a one-dimensional Brownian motion under $\mathbb{P}^*$, and $(\alpha,\mu,\gamma,\pi_0) \in \mathbb{R}_{*,+}^{4}$. As a result, the stochastic transition probability matrix from time $t$ to time $t+h$ can be recovered as follows:

\[
Q_{t}(t+h)=\exp\left(\int_{t}^{t+h} \pi_u \Lambda du\right) = M \exp\left( \left\{\int_{t}^{t+h} \pi_{u} du\right\} \log(D)\right) M^{-1}.
\]
In this framework, the logarithmic spread associated with the zero-coupon bond of maturity $t+h$ and rated as $R \in \mathcal{R}$ at time $t$ (this bond price is denoted by $P_{t}^{(R)}(t+h)$) is equal to $s_{t}^{(R)}(t+h)=-\frac{1}{h}\log\left(1 - L_{R} \mathbb{E}^{*}\left[Q^{R,R_d}_t(t+h) ~|~\mathcal{F}_t\right] \right)$ where $L_R$ denotes the loss-given-default (which is set to $0.75$ for all ratings in our numerical applications). We recover $P_{t}^{(R)}(t+h)$ using the following relation
\[
P_t^{(R)}(t+h)=P_t(t+h)\exp(-h \times s_{t}^{(R)}(t+h)).
\]

\section{Numerics}
\label{part2}

The numerical study is organized around three questions, corresponding to the three practical requirements that motivate the construction of a surrogate ALM model.

First, we assess whether the proposed signature-based approximation can accurately replicate the discounted cashflows generated by the ALM model. This is an in-distribution experiment: the surrogate is trained and tested under the same central ESG distribution. We evaluate both the pathwise approximation error, through the out-of-sample $R^2$, and the induced error on the VIF, which is the quantity of primary practical interest.

Second, still under this fixed distribution of economic scenarios, we compare the signature-based model with standard alternatives from functional data analysis and time series extrinsic regression. This benchmark should therefore be read as a comparison of surrogate model classes on the same learning problem, and not as a robustness analysis with respect to changes in the ESG distribution.

Third, we investigate whether a signature-based surrogate calibrated once under the central ESG distribution remains accurate when the distribution of economic scenarios is modified. To do so, we apply multiple perturbations to the parameters of the ESG and generate new datasets of scenarios under these new distributions, while the whole fitted
surrogate pipeline -- the reduction map $f_\theta$, the standardization parameters
and the regression coefficients -- is kept fixed. This experiment reflects a practical use case in which an insurer calibrates the surrogate model under a central ESG calibration and subsequently uses it after market conditions have changed, requiring the ESG to be recalibrated while avoiding a full retraining of the surrogate.

For this section, we use the ALM model and the ESG model specified in Section~\ref{ALM model}. We first rely on a central dataset composed of 2000 i.i.d. economic scenarios. For each scenario, we compute the associated discounted cashflows using \eqref{CF alm}. The dataset is split into a training sample containing $70\%$ of the observations and a test sample containing the remaining $30\%$. Unless stated otherwise, all in-distribution experiments are performed using this central ESG calibration.

All experiments were carried out on a single machine equipped with an Intel Core i7-1355U CPU (13th Gen, 1.70 GHz) and 16 GB of RAM. No GPU acceleration or parallelization was used. The generation of cashflows for the 2000 central scenarios using the full ALM model required approximately 3 hours of computation time. This cost is the main motivation for replacing repeated ALM evaluations by a calibrated surrogate whenever possible.

\subsection{In-distribution performance of signature-based models}
\label{subsection:exp_central}

This subsection addresses the first question stated above. We train and test the signature-based model under the central ESG distribution. The objective is to determine whether a finite set of signature features is sufficiently expressive to reproduce the pathwise ALM cashflows, and to identify a specification offering a satisfactory accuracy--time trade-off.

We consider several configurations of the approximation model introduced in Section~\ref{part1} in order to study the impact of the transformations applied to the input path, the type of regularization used, and the truncation order of the signature. The results are reported in Table~\ref{complete_sig}.

\begin{table}[H]
\centering
\footnotesize
\begin{tabular}{lccccc}
\toprule
Regularization & Order & Lead-lag & Cumulative & $R^2$ (\%) & Time (s) \\
\midrule

\multicolumn{6}{l}{\textit{Ridge regression}} \\

& \textbf{2} & $\times$ & $\times$ & 44.87 & 0.2 \\
&            & $\times$ & \checkmark & 78.17 & 0.2 \\
&            & \checkmark & $\times$ & 45.04 & 0.8 \\
&            & \checkmark & \checkmark & \textbf{80.44} & 0.7 \\
\addlinespace
& \textbf{3} & $\times$ & $\times$ & 80.13 & 2.7 \\
&            & $\times$ & \checkmark & 91.33 & 2.8 \\
&            & \checkmark & $\times$ & 78.23 & 5.9 \\
&            & \checkmark & \checkmark & \textbf{92.72} & 5.8 \\
\addlinespace
& \textbf{4} & $\times$ & $\times$ & 87.74 & 5.7 \\
&            & $\times$ & \checkmark & \textbf{94.63} & 7.1 \\
&            & \checkmark & $\times$ & 82.70 & 101.5 \\
&            & \checkmark & \checkmark & 94.48 & 89.2 \\
\addlinespace
& \textbf{5} & $\times$ & $\times$ & 86.48 & 48.1 \\
&            & $\times$ & \checkmark & \textbf{95.25} & 56.8 \\

\midrule

\multicolumn{6}{l}{\textit{Lasso regression}} \\

& \textbf{2} & $\times$ & $\times$ & 45.89 & 0.2 \\
&            & $\times$ & \checkmark & 78.31 & 0.2 \\
&            & \checkmark & $\times$ & 45.37 & 0.7 \\
&            & \checkmark & \checkmark & \textbf{78.37} & 0.6 \\
\addlinespace
& \textbf{3} & $\times$ & $\times$ & 80.77 & 34.9 \\
&            & $\times$ & \checkmark & 91.30 & 25.8 \\
&            & \checkmark & $\times$ & 80.53 & 313.3 \\
&            & \checkmark & \checkmark & \textbf{91.62} & 391.3 \\

\bottomrule
\end{tabular}
\caption{\textbf{Comparative results of signature specifications.}
Out-of-sample $R^2$ values and training times for different truncation orders, regularization methods and path transformations. The symbol $\checkmark$ (resp. $\times$) means that the transformation has been applied (resp. has not been applied). For each truncation order and each regularization strategy, the highest $R^2$ is highlighted in bold. For Ridge regression at truncation order $5$, the lead-lag specifications were not considered due to computational constraints. Truncation orders $4$ and $5$ were not considered for Lasso regression for the same reason. The cost of cross-validating $\lambda$ is included in the reported training times.}
\label{complete_sig}
\end{table}

The results show that signature-based regressions provide strong out-of-sample predictive performance under the central ESG distribution. Ridge and Lasso regularization lead to comparable accuracy levels, but Ridge regression is substantially faster, as its estimator is obtained through a closed-form formula. This makes Ridge the preferred specification in the present setting. The cumulative transformation has a clear positive impact on predictive performance, while adding no material storage or computational burden. By contrast, the lead-lag transformation does not provide a systematic improvement in $R^2$ and significantly increases the computational cost, since it doubles the path dimension and increases the number of signature features by a factor of order $2^N$, where $N$ is the truncation order. Although the Ridge model of order $5$ with cumulative transformation achieves the highest $R^2$ among the reported signature specifications, the Ridge model of order $4$ without lead-lag and with cumulative transformation provides a more attractive accuracy--time trade-off. We therefore use this specification for the graphical evaluation below. We denote by $\widehat{\text{CF}}$ the output of our signature-based approximation described in Figure \ref{fig:signature_model}.

\begin{figure}[H]
    \centering
    \subfloat[Scatter plot of $(\widehat{\text{CF}},\text{CF})$.]{
        \includegraphics[width=0.3\linewidth]{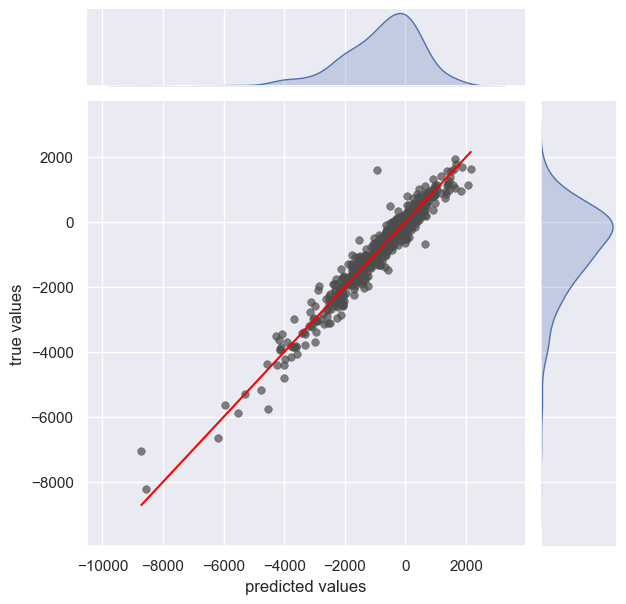}
    }\hfill
    \subfloat[Empirical distributions of $\widehat{\text{CF}}$ and CF.]{
        \includegraphics[width=0.3\linewidth]{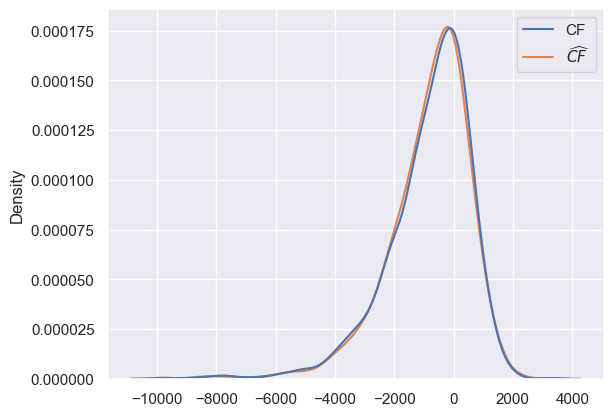}
    }\hfill
    \subfloat[QQ-plot between the empirical distributions of CF and $\widehat{\text{CF}}$.]{
        \includegraphics[width=0.3\linewidth]{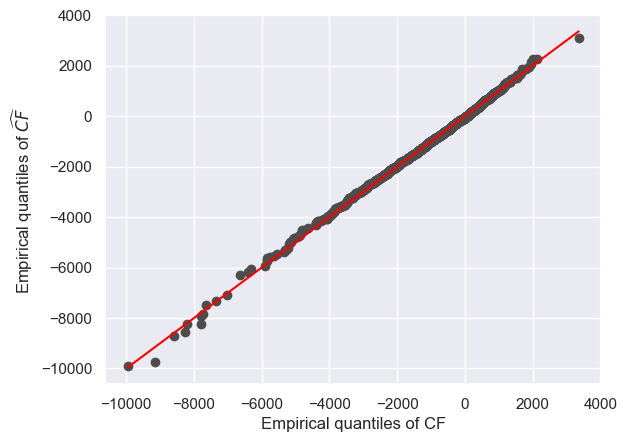}
    }

    \caption{%
      \textbf{Empirical study of cashflow replication.}
      These results are obtained with a truncation order equal to 4, without lead-lag transform, with cumulative transform and with a Ridge regularization.}
    \label{graph::distrib_CF}
\end{figure}

\begin{figure}[H]
    \centering
    \subfloat[Scatter plot of $(\widehat{\text{CF}},\varepsilon)$.]{%
         \includegraphics[width=0.3\linewidth]{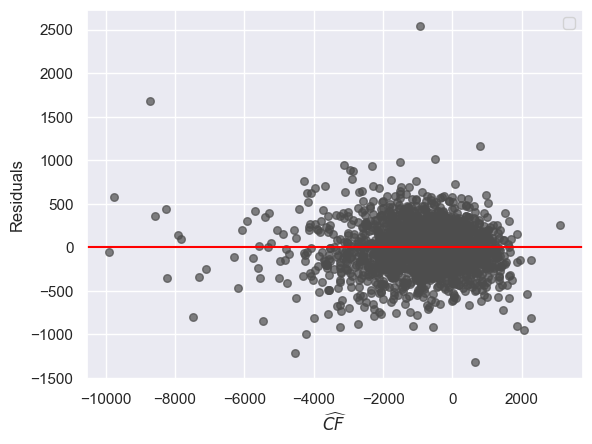}
    }\hfill
    \subfloat[QQ-plot between the empirical distribution of standardized residuals and a standard Gaussian distribution.]{%
         \includegraphics[width=0.25\linewidth]{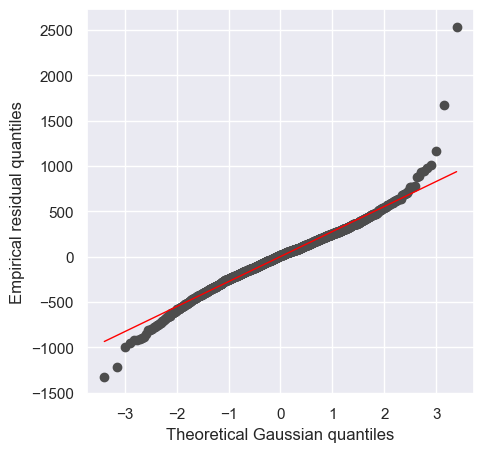}
    }\hfill
    \subfloat[Gaussian kernel estimate of the conditional variance of residuals.]{%
         \includegraphics[width=0.3\linewidth]{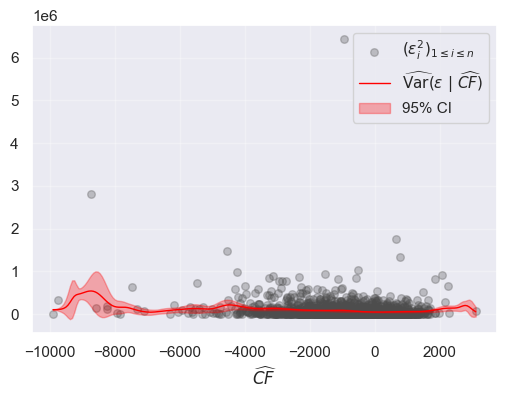}
    }
    
    \caption{\textbf{Empirical study of residuals.} These results are obtained with a truncation order equal to 4, without lead-lag transform, with cumulative transform and with a Ridge regularization.}
    \label{graph::distrib_residuals}
\end{figure}

While our performance metric focuses on the pathwise replication of cashflows through the approximation $\Phi_{\text{sig}} \simeq \Phi$, the resulting agreement between the empirical distributions of CF and $\widehat{\text{CF}}$ follows naturally from this strong form of replication, as illustrated in Figure \ref{graph::distrib_CF}(b) and Figure \ref{graph::distrib_CF}(c). We emphasize the strong ability of the signature-based model to reproduce the empirical distribution with high accuracy, including in the tails. While the $R^2$ metric provides a global assessment of the pathwise replication accuracy, it does not fully characterize the structure of the remaining approximation error. It is natural to analyze the residual $\varepsilon$ defined as $\varepsilon:= \Phi(X) - \Phi_{\text{sig}}(X)$ in order to assess whether the remaining error exhibits systematic patterns or regime-dependent behaviors. We provide in Figure \ref{graph::distrib_residuals}(a) the scatter plot of $(\widehat{\text{CF}},\varepsilon)$: it appears that there is no predictable structure of the residuals given the prediction of the model. We provide in Figure \ref{graph::distrib_residuals}(b) the QQ-plot between the empirical distribution of standardized residuals and a standard Gaussian distribution. Finally, we provide in Figure \ref{graph::distrib_residuals}(c) an estimate of $\text{Var}(\varepsilon ~|~ \widehat{\text{CF}})$ the conditional variance of the residual given the prediction of the model, using a Nadaraya–Watson estimator.

We recall that, in practice, the quantity of interest is the VIF, whose Monte-Carlo estimator is the empirical mean of the discounted profit and loss cashflows. Therefore, beyond the pathwise replication accuracy measured by the out-of-sample $R^2$, it is important to assess the error induced by the surrogate after aggregation over scenarios. When comparing the approximate VIF, computed as the empirical mean of the signature-based model outputs, with the VIF obtained from the full ALM model, we obtain a relative error of -0.20\%. This low relative error indicates that the residual approximation errors average out well. This is particularly relevant when the surrogate is used to increase the effective sample size for the Monte-Carlo estimator of the VIF. Once the surrogate has been calibrated and validated on a finite set of full ALM runs, additional economic scenarios can be evaluated at very low computational cost using $\Phi_{\mathrm{sig}}$. To illustrate this point, we generate an additional dataset of 10,000 i.i.d. economic scenarios drawn from the same law as the central dataset. The surrogate is not retrained: the model calibrated above -- with truncation order 4, cumulative transform,
no lead-lag and Ridge regularization -- is kept fixed and simply applied to these new
scenarios. It achieves an out-of-sample $R^2$ of 94.91\%, in line with the 94.63\%
reported in Table~\ref{complete_sig}.

We then quantify the error induced on the VIF at this larger scale. For each scenario
$k$, let $\varepsilon_k := \mathrm{CF}_k - \widehat{\mathrm{CF}}_k$ denote the deviation
between the surrogate output and the cashflow produced by the full ALM model. The error
induced by the surrogate on the VIF is exactly the empirical mean $\bar{\varepsilon}$ of
these deviations, for which the central limit theorem provides an asymptotic confidence
interval. Over the 10,000 scenarios, we obtain $\bar{\varepsilon} = -4.41$ with a 95\%
confidence interval equal to $[-10.64,1.81]$. Since this interval contains zero, the
systematic error induced by the surrogate on the VIF is not statistically detectable at
the resolution of a Monte-Carlo estimator based on 10,000 scenarios.

This experiment supports the use case described above: the surrogate is calibrated once
on a small number of full ALM runs and subsequently applied to a much larger set of
economic scenarios, thereby reducing the Monte-Carlo variance of the VIF estimator
without requiring a proportional increase in the number of expensive full ALM
evaluations.

\subsection{Benchmark comparison under the central ESG distribution}

We now compare the proposed signature-based approach with alternative surrogate models. All models in this subsection are trained and evaluated on the same central ESG dataset and on the same train-test split as above. Consequently, this benchmark assesses the relative predictive performance and computational cost of competing regression architectures under a fixed distribution of economic scenarios. It does not measure robustness to a change in the ESG distribution; this point is addressed separately in Section~\ref{sec:standard_formula_robustness}. All benchmark models are applied after the same preprocessing map $f_\theta$ as the signature-based models. In other words, the benchmark is not performed on the raw ESG output but on the common reduced path representation described in Section~\ref{part1}. This choice reflects the practical construction of surrogate models in high-dimensional ALM applications: dimension reduction is a shared preprocessing step, while the benchmark compares the regression architectures used after this step.

As a benchmark to assess the proposed signature-based approach, we consider methodologies stemming from functional data analysis (FDA) and time-series extrinsic regression (TSER). In the functional linear regression setting, the scalar response $Y$ is modeled as

\[
Y = \beta_0 + \sum_{j=1}^{d} \int_0^T \beta_j(t) X_t^{\word{j}} dt + \varepsilon,
\]
where the coefficient functions $\beta_j(t)$, $j=1,\cdots,d$ are estimated after projecting $(X_t)_{t \in [0,T]}$ onto a finite-dimensional basis such as B-splines, Fourier functions, or functional principal components. The comparison between signature-based models and functional linear regression has already been investigated in \citep{FERMANIAN2022105031} where signature-based models are shown to provide competitive or superior performance on several functional prediction tasks such as synthetic toy examples and a real-world air quality dataset. More generally, the TSER framework addresses the problem of predicting a scalar output from a path using a wide range of strategies, including basis representations, convolution-based models, distance-based models, and deep learning architectures. Functional data analysis (FDA) models are implemented following \citep{FERMANIAN2022105031} and we rely on the library \texttt{aeon} \citep{middlehurst2024aeon} which offers a unified implementation of TSER methods. We provide comparative results in Table \ref{tab:deep_ensemble}.

\begin{table}[H]
\centering
\footnotesize
\begin{tabular}{lcc}
\toprule
Method & $R^2$ (\%) & Time (s) \\
\midrule
\midrule
\multicolumn{3}{l}{\textit{Functional linear regression}} \\
B-spline & 73.28 & 0.2 \\
Fourier & 68.48 & 0.4 \\
FPCA & 33.14 & 0.2 \\
\midrule
\multicolumn{3}{l}{\textit{Convolution-based methods}} \\
RocketRegressor &  46.03 & 68.01 \\
MiniRocketRegressor &  69.32 & 2.3 \\
MultiRocketRegressor &  63.98 & 16.2 \\
HydraRegressor & 43.41 & 3.6 \\
MultiRocketHydraRegressor & 62.02 & 47.6 \\
\midrule
\multicolumn{3}{l}{\textit{Interval-based methods}} \\
RandomIntervalRegressor & 71.15 & 128.9 \\
IntervalForestRegressor & 66.89 & 57.3 \\
CanonicalIntervalForestRegressor & 45.89 & 390.8 \\
DrCIFRegressor & 36.70 & 794.8 \\
QUANTRegressor & 60.29 & 46.0 \\
RandomIntervalSpectralEnsembleRegressor & 8.45 & 96.2 \\
TimeSeriesForestRegressor & 65.38 & 61.8 \\

\midrule
\multicolumn{3}{l}{\textit{Distance-based methods}} \\
KNeighborsTimeSeriesRegressor & 30.03 & 68.5 \\

\midrule
\multicolumn{3}{l}{\textit{Feature-based methods}} \\
Catch22Regressor & 67.21 & 52.1 \\
SummaryRegressor & 66.44 & 19.3 \\
TSFreshRegressor & 21.69 & 369.7 \\
FreshPRINCERegressor & 86.82 & 19224.5 \\

\midrule
\multicolumn{3}{l}{\textit{Deep learning architectures}} \\
TimeCNNRegressor & 84.72 & 2025.7 \\
ResNetRegressor & 87.56 & 2568.2 \\
IndividualInceptionRegressor & \textbf{94.70} & 4663.6 \\
IndividualLITERegressor & \textbf{94.00} & 1850.2 \\
LITETimeRegressor & \textbf{96.38} & 9000.1 \\
FCNRegressor & 79.42 & 6698.34 \\
EncoderRegressor & * & 1260 \\
MLPRegressor & 19.37 & 1026.0 \\
RecurrentRegressor & 35.78 & 2421.0 \\
DisjointCNNRegressor & 69.59 & 22539.0 \\
\bottomrule
\end{tabular}
\caption{\textbf{Comparison across linear functional regression and TSER models.}
Out-of-sample $R^2$ values and training times for the different models. All models are trained on the same dimension-reduced economic scenarios obtained through the preprocessing map $f_\theta$. The three highest $R^2$ values are highlighted in bold. Negative $R^2$ values are omitted and denoted by $*$.}
\label{tab:deep_ensemble}
\end{table}

\begin{figure}[H]
    \centering
    \includegraphics[width=0.8\linewidth]{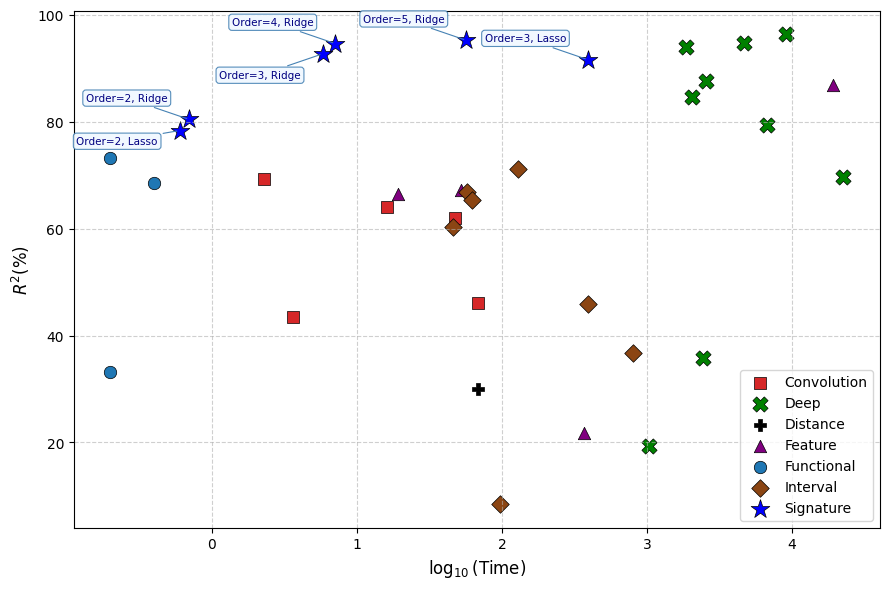}
    \caption{\textbf{Predictive performance versus computational cost across methods under the central ESG distribution.}
Each point represents a model, with the base-10 logarithm of training time (in seconds) on the horizontal axis and the out-of-sample $R^2$ (in \%) on the vertical axis.
Signature-based models (Ridge and Lasso regularization, truncation orders $N \in \{2, 3, 4, 5\}$) correspond to the best-performing specifications reported in Table~\ref{complete_sig} for each $(N, \text{regularization})$ pair.
All competing methods listed in Table~\ref{tab:deep_ensemble} are included for comparison.
All models are trained and tested under the central ESG distribution; the figure therefore compares model classes under a fixed law of economic scenarios.
}
    \label{fig:benchmark}
\end{figure}

Deep learning architectures achieve high predictive performance but at the cost of substantial training times. In contrast, feature-based and interval-based methods provide moderate accuracy with lower computational burden. To better illustrate the trade-off between predictive performance and computational cost across methods, we provide a visual comparison in Figure \ref{fig:benchmark}.

\subsection{Robustness to ESG distribution shifts}
\label{sec:standard_formula_robustness}

\begin{figure}[H]
    \centering
    \subfloat[Boxplot of $R^2$.]{
        \includegraphics[width=0.3\linewidth]{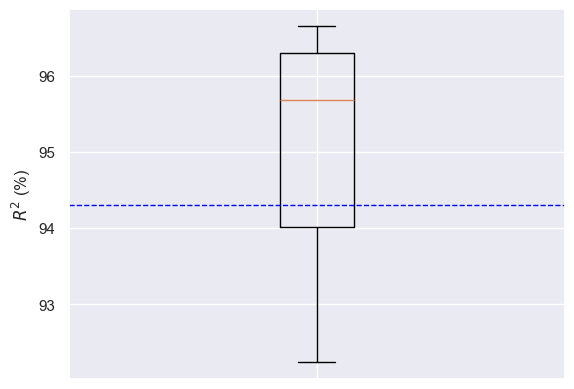}
    }\hfill
    \subfloat[Boxplot of $\mathrm{RE}_{\mathrm{VIF}}$.]{
        \includegraphics[width=0.3\linewidth]{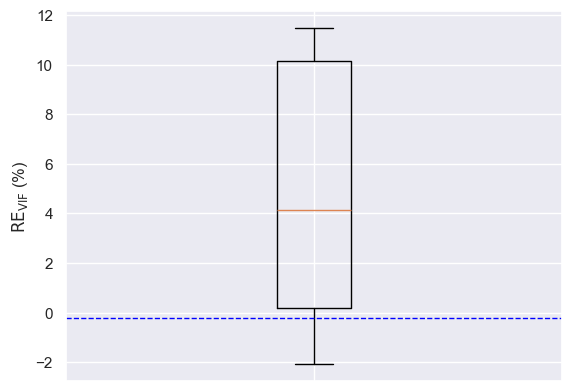}
    }\hfill
    \subfloat[Boxplot of $\mathrm{BSE}_{\mathrm{VIF}}$.]{
        \includegraphics[width=0.3\linewidth]{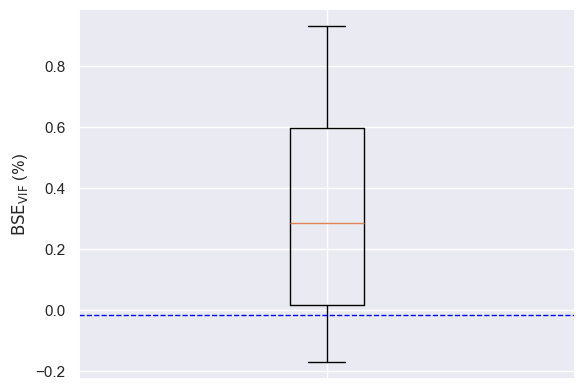}
    }
    \caption{\textbf{Robustness under ESG parameter perturbations.}
The boxplots report the distribution of out-of-sample $R^2$ values, VIF relative errors and balance-sheet normalized errors across $M_Z=50$ perturbed ESG parameterizations. For each parameterization, $M=1000$ new economic scenarios are generated and evaluated with the full ALM model. The signature-based surrogate is kept fixed after calibration on the central ESG distribution. The dotted horizontal line corresponds to the value obtained under the central ESG distribution on the reference dataset of $600$ scenarios.}
\label{fig:robustness_distribution}
\end{figure}

The previous subsections are conditional on the central ESG distribution. However, the regression coefficients of the signature-based surrogate are learned from scenarios generated under a given ESG calibration, and therefore depend on the distribution of the input paths. Strong in-distribution results are thus not sufficient to conclude that the surrogate remains accurate after a change in the law of economic scenarios. We address this issue through a controlled out-of-distribution experiment. Let $Z_{\rm f,0}$ denote the central market environment and let $p_0 := \mathcal{L}(X(Z_{\rm f, 0}))$ be the associated central ESG distribution. By a slight abuse of notation, we assume that $Z_{\rm f,0} \in \mathbb{R}^p$ denotes the set of parameters of the ESG under this central market environment. The signature-based surrogate $\Phi_{\mathrm{sig}}^{(0)}$ is trained once under $p_0$, using the central dataset described above. For $\delta>0$, we consider the random variable $Z_{\rm f,\delta} =Z_{\rm f,0}\odot |\mathcal{E}|_{p}$ with $\mathcal{E} \sim \mathcal{N}(\mathbf{1}_p, \delta^2 I_{p})$ independent of the ESG Brownian noise, $\mathbf{1}_p=(1,\cdots,1) \in \mathbb{R}^p$, and $|(x_1,\cdots,x_p)|_p:=(|x_1|,\cdots,|x_p|)$. This method ensures that the signs of the parameters do not change after the perturbation. We sample $M_{Z}=50$ i.i.d. samples of $Z_{\rm f,\delta}$, denoted by $(z_i)_{1 \le i \le M_{Z}}$, and for each $z_i$, we generate a new sample of $M=1000$ economic scenarios $(X_k^{(z_i)})_{1 \leq k \leq M}$ under $\mathcal{L}(X(z_i))$. Then we run the full ALM model to obtain the reference cashflows
\[
\text{CF}_k^{(z_i)} = \Phi(X_k^{(z_i)}),
\]
and compare them with the surrogate predictions
\[
\widehat{\text{CF}}_k^{(z_i)} = \Phi_{\mathrm{sig}}^{(0)}(X_k^{(z_i)}).
\]
No retraining is performed after the shock: the PCA projection, standardization parameters and regression coefficients learned under the central ESG distribution are kept fixed. For each perturbed parameter vector $z_i$, we compute the out-of-sample $R^2$ on the corresponding newly generated dataset, together with the relative error on the VIF and the balance-sheet normalized error
\[
\mathrm{RE}_{\mathrm{VIF}}(z_i)
=
\frac{
\widehat{\mathrm{VIF}}(z_i)-\mathrm{VIF}(z_i)
}{
|\mathrm{VIF}(z_i)|
}, \quad \mathrm{BSE}_{\mathrm{VIF}}(z_i)=\frac{
\widehat{\mathrm{VIF}}(z_i)-\mathrm{VIF}(z_i)
}{
\mathrm{BS}
},
\]
where
\[
\mathrm{VIF}(z_i)
=
\frac{1}{M}\sum_{k=1}^M \text{CF}_k^{(z_i)},
\qquad
\widehat{\mathrm{VIF}}(z_i)
=
\frac{1}{M}\sum_{k=1}^M \widehat{\text{CF}}_k^{(z_i)},
\]
and BS denotes the total balance-sheet value. In the ALM model considered here, there is no explicit own-funds account. Hence the total balance-sheet size is fixed by the initial liability portfolio and does not depend on the perturbed ESG parameterization $z_i$. While $\mathrm{RE}_{\mathrm{VIF}}$ measures the error relative to the target VIF itself, $\mathrm{BSE}_{\mathrm{VIF}}$ measures the economic materiality of the approximation error relative to the overall scale of the balance sheet. This additional metric is useful when the VIF is small, since a moderate absolute error may lead to a large relative error while remaining immaterial at the balance-sheet level.
The resulting distributions of $R^2$, $\mathrm{RE}_{\mathrm{VIF}}$ and $\mathrm{BSE}_{\mathrm{VIF}}$ are reported in Figure~\ref{fig:robustness_distribution}. Due to computational constraints, each perturbed ESG calibration is evaluated on $M=1000$ scenarios instead of the $2000$ scenarios used in the central experiment (see Section \ref{subsection:exp_central}). To facilitate comparison with the in-distribution benchmark, the dotted horizontal line represents the value obtained on the central reference test dataset of $600$ scenarios.

Overall, the results suggest that the surrogate remains accurate under moderate perturbations of the ESG parameterization. Although the regression coefficients are learned under the central ESG distribution, the signature features appear to capture stable properties of the ALM cashflow functional that remain informative beyond the training distribution. This provides empirical evidence that the method can be used for fast sensitivity analyses under changes in market assumptions, provided that the perturbed ESG distributions remain close enough to the central calibration.

\section{Conclusion and further work}
\label{conclusion}

In this paper, we introduced the use of the signature in order to construct a tractable and robust approximation model of the sum of discounted cashflows produced by a realistic ALM model. The proposed approach relies on a linear regression on the time-augmented signature components of the economic scenarios provided as inputs. 

The numerical results show that the proposed method achieves strong out-of-sample replication accuracy under the central ESG distribution. In particular, low-order signature models already provide high predictive performance, while remaining computationally efficient. The comparison with functional regression and time-series regression benchmarks further shows that signature-based models offer a competitive accuracy--time trade-off. Unlike more computationally intensive architectures, the proposed estimator remains easy to calibrate, since it is ultimately fitted using regularized linear regression.

Beyond pathwise replication accuracy, the results also show that the surrogate provides accurate estimates of balance-sheet quantities such as the VIF. Since the VIF is obtained by averaging discounted cashflows across economic scenarios, the relevant operational metric is not only the out-of-sample $R^2$, but also the error induced after aggregation. In our experiments, the relative error on the VIF remains small, suggesting that the residual approximation errors average out well over all economic scenarios.

We also investigated the robustness of the surrogate under changes in the distribution of economic scenarios. The results obtained under perturbations of the ESG parameterization indicate that the surrogate remains accurate when considering perturbed parameters of the ESG specification. This provides empirical evidence that the signature representation captures stable features of the ALM cashflow functional, and can therefore be useful for fast sensitivity analyses under changes in market assumptions.

The proposed approach is not intended to replace the full ALM model for regulatory production. Instead, it should be viewed as a complementary tool that can reduce computational costs in exploratory studies, large-scale sensitivity analyses, nested simulation settings, and sandbox experiments. Once calibrated and validated, the surrogate can also be used to evaluate additional economic scenarios at low cost, thereby increasing the effective sample size available for estimating quantities such as the VIF.

Several directions remain open for future work. First, the robustness analysis could be extended to broader changes in ESG model families, more severe market stresses, and changes in non-financial assumptions such as lapses, mortality, or management rules. Second, the methodology could be tested on larger and more realistic ALM models, including richer liability portfolios and additional asset classes. Finally, it would be valuable to develop uncertainty quantification procedures for the surrogate error, in order to provide confidence bounds on the induced error for balance-sheet quantities. 

\mbox{}
\clearpage
\bibliographystyle{abbrvnat}
\bibliography{ref}

\newpage

\appendix
\renewcommand{\thesection}{Appendix \Alph{section}}

\section{Path transformations}
\label{appendix A}

In this appendix, we describe the main path transformations used in this article prior to the computation of the truncated time-augmented signature. Such transformations are commonly applied in the signature literature to enhance its  expressiveness \citep{chevyrev2016primer,levin2016learningpastpredictingstatistics,gyurko2013extracting,andres2024signature,lyons2014feature}.

\paragraph{Lead-lag transform.}

Given a discrete-time path taking values in $\mathbb{R}^d$, the lead-lag transform constructs a new path taking values in $\mathbb{R}^{2d}$ defined by duplicating each coordinate with a slight temporal offset. More precisely, for a stochastic process $X$ taking values in $\mathbb{R}^d$ and a partition $0=t_0<t_{1 / 2}<t_1<\cdots<t_{K-1 / 2}<t_K=T$, the lead-lag transformation of $X$ on the partition $\left(t_{i / 2}\right)_{i=0, \ldots, 2K}$ is the $2d$-dimensional path $t \mapsto\left(X_t^{\text {lead }}, X_t^{\text {lag }}\right)$ defined on $[0, T]$ where:

\begin{enumerate}
    \item the lead process $t \mapsto X_t^{\text {lead }}$ is the linear interpolation of the points $\left(X_{t_{i / 2}}^{\text {lead }}\right)_{i=0, \ldots, 2 K}$ with:

\[
X_{t_{i / 2}}^{\text {lead }}= \begin{cases}X_{t_j} & \text { if } i=2 j \\ X_{t_{j+1}} & \text { if } i=2 j+1\end{cases}
\]

\item the lag process $t \mapsto X_t^{\text {lag }}$ is the linear interpolation of the points $\left(X_{t_{i / 2}}^{\text {lag }}\right)_{i=0, \ldots, 2 K}$ with:

\[
X_{t_{i / 2}}^{\text{lag}}= \begin{cases}X_{t_j} & \text { if } i=2 j \\ X_{t_j} & \text { if } i=2 j+1\end{cases}
\]
\end{enumerate}

An important property of the lead-lag transform is that it allows one to capture the quadratic variation of a path in the signature components. More precisely, the Lévy area of the lead-lag transform is the quadratic variation up to a factor $\frac{1}{2}$ as stated by the following proposition.

\begin{proposition}[see for instance \citep{chevyrev2016primer}]
Let $t_0=0<t_1<\cdots<t_K=T$ be a partition of $[0, T]$ and $\left(X_{t_k}\right)_{k=0, \cdots, K}$ be the vector of observations of a real-valued process $X$ on this partition. The Lévy area of the lead-lag transformation of $\left(X_{t_k}\right)_{k=0, \cdots, K}$ is equal to the quadratic variation of $X$ on the partition $\left(t_k\right)_{k=0, \cdots, K}$ up to a factor $\frac{1}{2}$, i.e.
\[
\frac{1}{2}\left(\int_0^T\left(X_t^{\text {lead }}-X_0^{\text {lead }}\right) d X_t^{\text {lag }}-\int_0^T\left(X_t^{\text {lag }}-X_0^{\text {lag }}\right) d X_t^{\text {lead }}\right)=\frac{1}{2} \sum_{k=0}^{K-1}\left(X_{t_{k+1}}-X_{t_k}\right)^2 .
\]   
\end{proposition}

\begin{figure}[H]
    \centering
    \subfloat[Lead and lag paths.]{
        \includegraphics[width=0.45\linewidth]{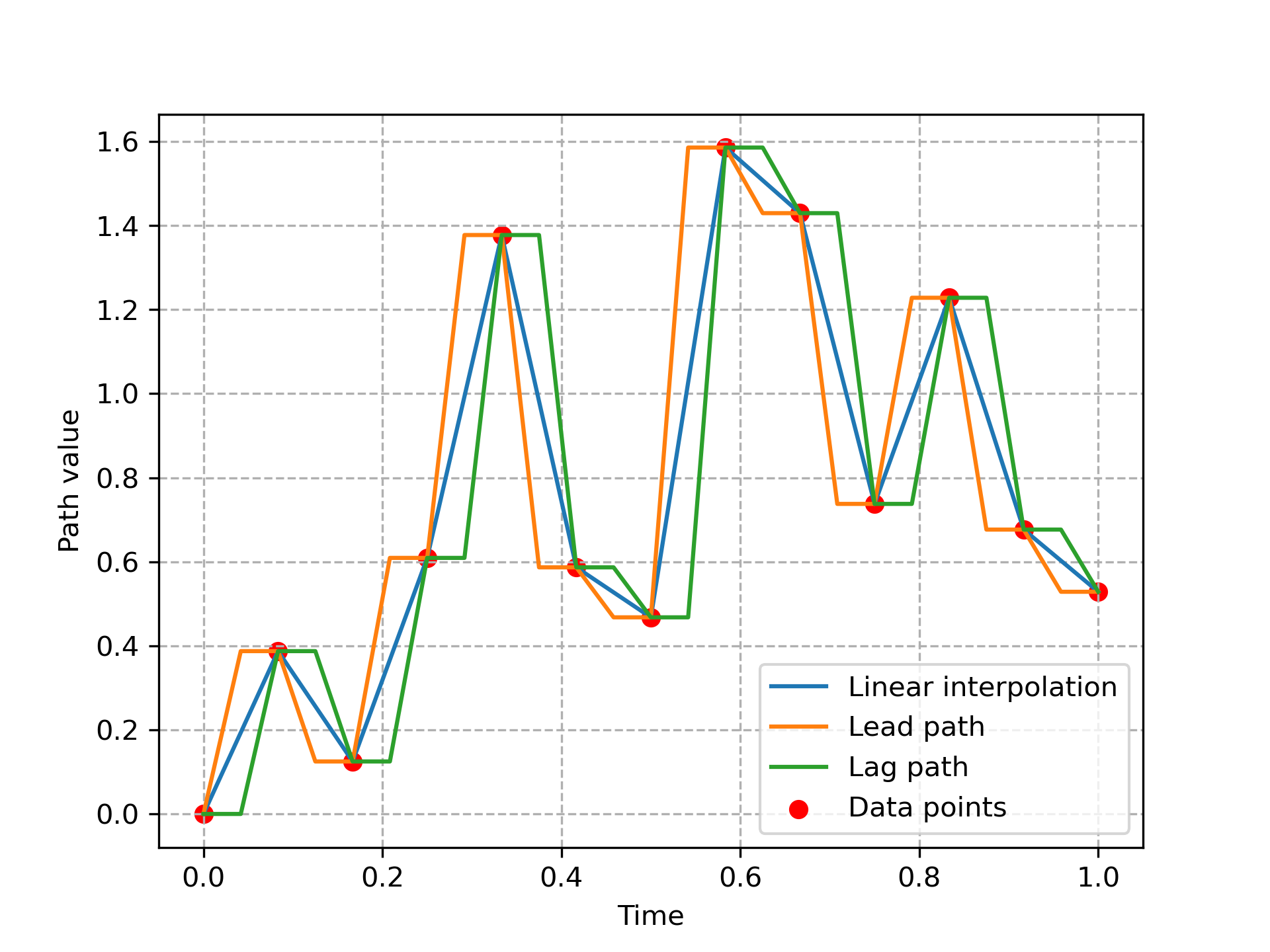}
    }\hfill
    \subfloat[Lead-lag transform.]{
        \includegraphics[width=0.45\linewidth]{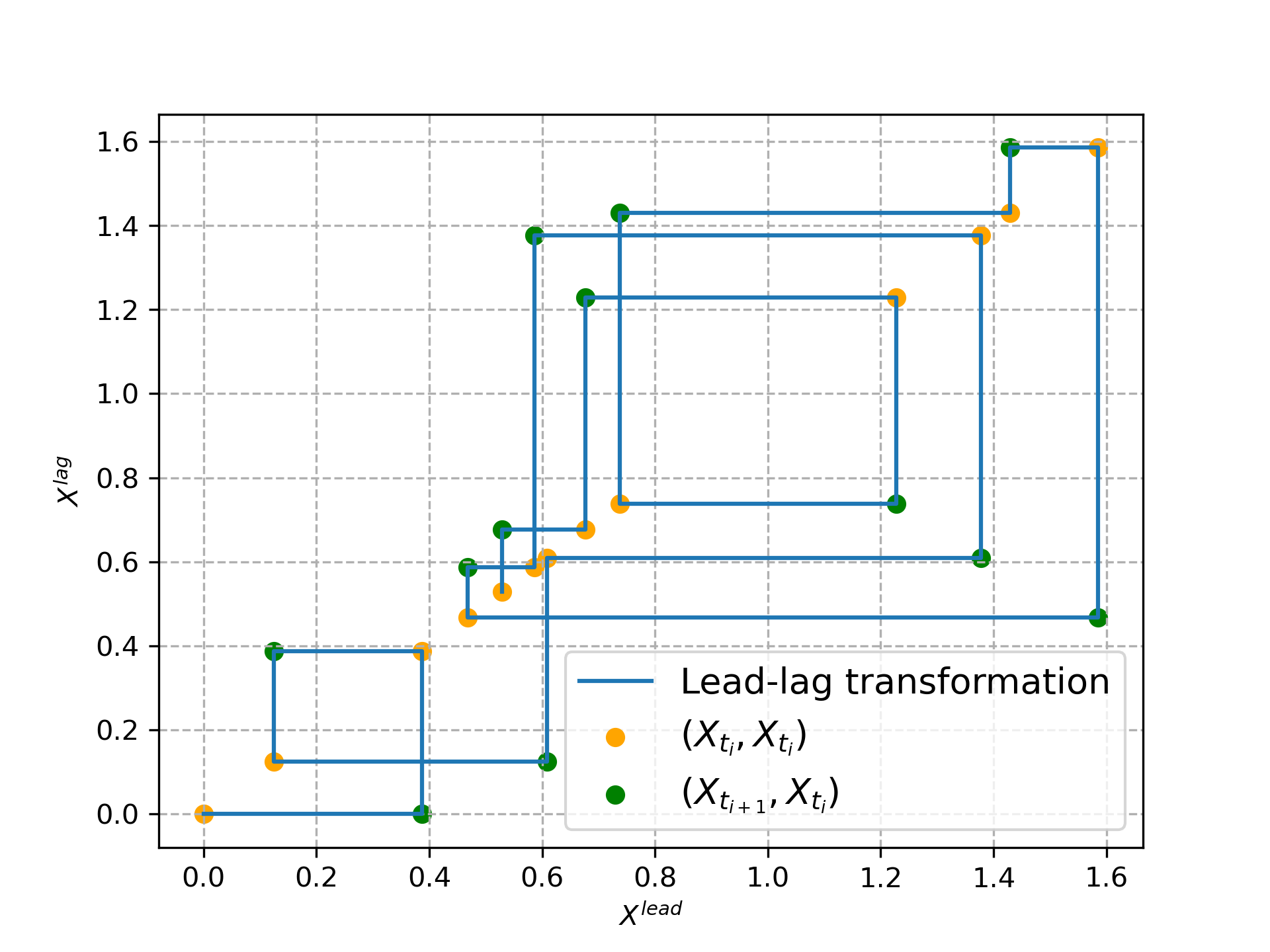}
    }

    \caption{%
      Lead-lag illustrations from \citep{andres2024signature}.}
    \label{leadlag}
\end{figure}

We provide in Figure \ref{leadlag} illustrations of the lead and lag paths, as well as the lead-lag transform. The main issue with the lead-lag transform is that for a $d$-dimensional path the resulting transformed path is now $2d$-dimensional and the number of suffix components of the time-augmented signature will increase by a factor $(\frac{2d+1}{d+1})^N \simeq 2^N$.

\paragraph{Cumulative transform.}

The cumulative transform consists of replacing the original discretized path $(X_{t_k})_{k=0,\cdots,K}$ by its cumulative version defined as follows for all $k=0,\cdots,K$:
\[
X_{t_{k}}^{\text{cumulative}}=\sum_{j=0}^{k} X_{t_j}.
\]
In other words, each component of the transformed path records the cumulative sum of the past values of the path. The cumulative transform was studied in \citep{chevyrev2016primer,andres2024signature}, where it was combined with the lead-lag transform in order to improve the predictive power of signature-based features. Indeed, when the cumulative transform is combined with the lead-lag transform, the signature components of the transformed path can be related to the statistical moments of the original path \citep{andres2024signature}[Remark A.1]. Nevertheless, we apply the cumulative transform even when it is used independently of the lead-lag transform. An important property of the cumulative transform is that it preserves the dimension of the underlying path, in contrast with the lead-lag transform which doubles the dimension of the path.

\section{ALM model specification}
\label{appendix B}

We summarize below the parameterization of the ALM model, including allocations, profit-sharing parameters, and portfolio structure.

\begin{table}[H]
    \centering
    \begin{tabular}{l | l | c}
        \toprule
        \textbf{Parameter} & \textbf{Description} & \textbf{Value} \\
        \midrule
        $r^G$ & guaranteed minimum rate & 0.015 \\
        $\pi_{pr}$ & participation rate & 0.9 \\
        $\rho^*$ & smoothing factor in the PSR dynamics &0.5 \\
        $n_{\text{basket}}$ & number of ZC bonds of each rating in the portfolio & 20 \\
        $w_e$ & equity allocation weight & 0.4 \\
        $w_b$ & risk-free bonds allocation weight & 0.4 \\
        $w_{b,\rm AAA}$ & AAA-rated bonds allocation weight & 0\\
        $w_{b,\rm AA}$ & AA-rated bonds allocation weight & 0.2\\
        $w_{b,\rm A}$ & A-rated bonds allocation weight & 0\\
        $w_{b,\rm BBB}$ & BBB-rated bonds allocation weight &0 \\
        $w_{b,\rm BB}$ & BB-rated bonds allocation weight & 0\\
        $w_{b,\rm B}$ & B-rated bonds allocation weight & 0\\
        $w_{b,\rm CCC}$ & CCC-rated bonds allocation weight &0 \\ 
        \bottomrule
    \end{tabular}
\caption{Parameterization of the ALM model.}
\label{tab:parameters}
\end{table}

\section{ESG specification}
\label{appendix C}

We provide a concise description of the models used to generate the economic scenarios. For each risk factor, we recall the model dynamics and the associated parameterization.

\subsection*{C.1 Risk-Free Rate Curve}

We provide in Table \ref{tab:parameters} the parameterization of the DDLMM model.

\begin{table}[H]
    \centering
    \begin{tabular}{l | c }
        \toprule
        Parameter & \text{Value} \\
        \midrule
        $\delta$ & 0.1\\ 
        $\kappa$ & 0.28 \\
        $\theta$ & 0.62 \\
        $a$ & 0.07 \\
        $b$ & 0.008 \\
        $c$ & 0.08 \\
        $d$ & 0.01 \\
        \bottomrule
    \end{tabular}
\caption{\textbf{DDLMM model parameterization.}}
\label{tab:parameters}
\end{table}

\subsection*{C.2 Credit Risk Model}

We provide in Table \ref{tab:parameters_JLT} the parameterization of the JLT model.

\begin{table}[H]
    \centering
    \begin{tabular}{l | c }
        \toprule
        Parameter & Value \\
        \midrule
        $\pi_0$ &3.48 \\ 
        $\alpha$ & 0.22 \\
        $\mu$ & 1.52 \\
        $\gamma$ & 0.82 \\
        \bottomrule
    \end{tabular}
\caption{\textbf{JLT model parameterization.}}
\label{tab:parameters_JLT}
\end{table}

\end{document}